\documentclass[letters]{IEEEtran}

\usepackage[utf8]{inputenc}
\usepackage{graphics,graphicx,color,epsfig}
\usepackage{amsmath,amssymb,amsthm,amsfonts,bm,bbm}
\usepackage{algorithmicx,algorithm,algpseudocode}
\usepackage{multirow}
\usepackage{cite}
\usepackage[normalem]{ulem}

\usepackage{tikz}
\usetikzlibrary{backgrounds,positioning,arrows,shapes.multipart,fit}
\usepackage{pgfplots}
\usepackage{pgfplotstable}
\usepgfplotslibrary{fillbetween}

\definecolor{dark2_0}{HTML}{1B9E77}
\definecolor{dark2_1}{HTML}{D95F02}
\definecolor{dark2_2}{HTML}{7570B3}
\definecolor{dark2_3}{HTML}{E7298A}
\definecolor{dark2_4}{HTML}{66A61E}
\definecolor{dark2_5}{HTML}{E6AB02}
\definecolor{dark2_6}{HTML}{A6761D}

\definecolor{urban_0}{HTML}{FEECB1}
\definecolor{urban_1}{HTML}{9C9B47}
\definecolor{urban_2}{HTML}{DD7F0C}
\definecolor{urban_3}{HTML}{445659}
\definecolor{urban_4}{HTML}{445511}

\definecolor{color_ae}{HTML}{1B9E77}
\definecolor{color_vae_awgn}{HTML}{D95F02}
\definecolor{color_vae_rbf}{HTML}{7570B3}
\definecolor{color_qam}{HTML}{E7298A}
\definecolor{color_agrell}{HTML}{66A61E}
\definecolor{color_vae_cau}{HTML}{D95F02}
\definecolor{color_vae_lap}{HTML}{D95F02}

\usepackage{subcaption}
\usepackage[textsize=tiny]{todonotes}
\presetkeys{todonotes}{color=blue!20}{}

\title{Design of Communication Systems using Deep Learning: A Variational Inference Perspective}
\author{Vishnu Raj \hspace{16pt} Sheetal Kalyani\\
   \hspace{-0 cm}Department of Electrical Engineering, \\ Indian Institute of Technology Madras, \\
   Chennai, India, 600 036. \\
   \texttt{\{ee14d213,skalyani\}@ee.iitm.ac.in}
}

\begin{document}
	\maketitle
	\begin{abstract}
		Recent research in the design of end to end communication system using deep learning has 
produced models which can outperform traditional communication schemes. Most
of these architectures leveraged autoencoders to design the encoder at the transmitter
and decoder at the receiver and train them jointly by modeling transmit symbols
as latent codes from the encoder. However, in communication systems, the receiver
has to work with noise corrupted versions of transmit symbols. 
Traditional autoencoders are not
designed to work with latent codes corrupted with noise.
In this work, we provide a framework to design end to end communication
systems which accounts for the existence of noise corrupted transmit symbols.
The proposed method uses deep neural architecture. An objective 
function for optimizing these models is derived based on the concepts of
variational  inference. Further, domain knowledge such as channel type can be
systematically integrated into the objective.
Through numerical simulation, the proposed method is shown to consistently produce 
models with better packing density and achieving it faster in multiple popular
channel models as compared to the
previous works leveraging deep learning models.
	\end{abstract}
	
	\section{Introduction}
The aim of any communication system is to perfectly reproduce the message
at the receiver sent by a 
transmitter through a channel between the sender and receiver. Due 
to the noise characteristics of the channel, the transmitted signal can get corrupted, 
and the exact reconstruction of the message may not happen at the receiver. 
A robust communication
system should be able to handle these corruptions due to the channel
and reproduce the message with maximum faithfulness at the receiver.

Traditional communication systems follow a block by block design, optimized within the block
for maximal performance. However, such a system may not result in a globally optimum 
solution across all blocks. The complexity of the signaling systems, 
along with the unknown effect from the channel, makes it difficult to design an optimal
system across all the blocks. Lately, deep learning has seen extraordinary success
in learning complex tasks involving natural signals such as images, speech, etc.
In the area of communication systems also, applications of deep learning have resulted
in improved results. 
In \cite{samuel2019learning}, a deep learning based approach by unfolding the projected
gradient descent algorithm is proposed for MIMO detection. A deep learning based
method for channel estimation in OFDM systems with one-bit quantization is developed
in \cite{balevi2019one}. Interestingly, the one-bit quantized OFDM systems with deep
learning based estimation is able to provide lower error than least-squares channel
estimation with unquantized samples. Interested readers are redirected to \cite{qin2019deep}
for a broad discussion on how deep learning can help to improve physical layer of
communication systems.

    \begin{table*}[!h]
        \centering
        \caption{Comparison of Proposed method to AE based models} \label{tab:characteristics}
        \begin{tabular}{l|c|c}
            \hline
            Characteristic & AE-based methods \cite{o2017introduction,dorner2018deep,aoudia2018model} & Proposed Method \\
            \hline
            \hline
            Basic Concept & Autoencoders & Variational Autoencoders \\
            Accounts for noise in latent code & No & Yes \\
            Constant SNR required at training & Yes & No \\
            Method for Power control & Through normalization layer & Through KL-divergence term in loss function\\
            Type of power constraint & Hard constraint & Soft constraint \\
            \hline
        \end{tabular}
    \end{table*}

In \cite{o2017introduction}, the authors proposed the fascinating
idea of an end to end design communication system based on the principles of 
autoencoders\cite{Ballard1987ModularLI}. 
However, to train the system end to end, channel knowledge was required for 
computing the weight updates during backpropagation. To overcome the problem of 
unknown channel model, \cite{dorner2018deep} proposed
to train the network in two phases: in the first phase train both the transmitter and
receiver networks in simulation with known channel model and second phase deploy 
the network in actual
channel and fine-tune the receiver network alone. A practical approach to train
systems from end to end without any assumptions about the channel is proposed in
\cite{raj2018backpropagating} based on simultaneous perturbation stochastic approximations.
Another method is proposed in \cite{aoudia2018model} based on output perturbations
at the transmitter. Approaches to approximate the channel distribution with
neural networks and use this as a surrogate channel for backpropagation are proposed
in \cite{o2018physical,ye2018channel}.

The success of deep learning approach for transceiver design
is not just limited to wireless communication systems. In the context of optical communication
systems, \cite{karanov2018end} introduces an end to end deep learning based optical
communication transceiver for generating robust transmit waveforms used for communication
which is achieved by using a modified ReLU activation at the output of transmitter. 
In molecular communication systems, a deep learning based approach to optimize the receiver
design in the presence of Inter-Symbol Interference (ISI) is presented in \cite{qian2019molecular}.
In the context of Underwater Acoustic communications, \cite{jiang2019deep} proposes a
novel channel estimation technique for Orthogonal Frequency Division Multiplexing (OFDM)
systems which is capable of providing better performance than traditional Least Squares (LS)
and Minimum Mean Square Error (MMSE) estimators.

Previous works on end to end communication system design using deep learning 
\cite{o2016learning,o2017introduction,dorner2018deep,raj2018backpropagating,aoudia2018model}
relied on Autoencoders (AE) for designing the encoder and 
decoder. One of the original purposes of AE is to perform dimensionality reduction \cite{hinton1994autoencoders}
by using the latent codes produced by the encoder as compressed representation.
The works in \cite{o2017introduction,dorner2018deep,aoudia2018model} used the concepts
of AE to train an encoder for mapping a symbol to be transmitted to a constellation point and a decoder for
decoding the learned mapping. However, when using AEs for end to end communication system design, two fundamental
problems remain.
\begin{enumerate}
    \item By using a normalization layer at the end of transmitter (encoder), the AE based designs effectively
    hard constrain the parameter space. The normalization layer was introduced to achieve
    power constraint at the encoder since otherwise, one can trivially increase the transmit power to
    achieve better reconstruction at the receiver. However, such a hard constraint in one of the layers of a 
    deep network will impact the loss surface and parameter space one can explore \cite{marquez2017imposing}. 
    This could lead to trading off
    better designs for hard power constraints.
    \item In the context of communication systems, the decoder has to operate at a noisy version of the latent
    code produced by the encoder (transmitter and channel combined). 
    However, autoencoders are not designed to act on noisy latent codes and to the best of our
    knowledge, there exists no theoretical work on the behavior of AEs in the presence of noisy
    latent codes.
    A variant of AEs known as \textit{Noisy Autoencoders} can be used to work with noisy inputs 
    \cite{vincent2008extracting}, but not with a noise corrupted latent variable.
\end{enumerate}
This motivated us to investigate models that can handle noisy latent codes, has a theoretical
backing for the same, and which also imposes power constraint but as a soft constraint - hence enabling
more exploration and subsequently leading to better constellation designs when compared to using AEs. 

We propose a method based on the principles of Variational Autoencoders (VAEs) 
\cite{kingma2013auto} which allows to account for the noisy latent codes and provides 
a systematic way for introducing soft constraints on transmit power. VAEs were originally 
proposed as a distribution approximating method for generative modeling. VAEs approximate 
a complete distribution of the latent codes, typically using a multivariate Gaussian 
distribution by characterizing the mean and variance of the distribution. The proposed 
method uses the encoder to predict only the location of the transmit symbol, and channel 
is the entity that adds corruption to the transmitted symbol. Hence the mean of 
the conditional distribution at receiver is decided by the encoder while the variance 
is dictated by the channel. Compared to the autoencoder based design in existing 
literature, this approach provides a new interpretation for the noise corruption 
happening to the transmitted symbols. 

The proposed approach and the new interpretation following it help in deriving 
objective functions which can include prior information about the channel models 
or domain-specific information and can be used to train transmitter and 
receiver jointly. We show that the models trained with loss functions derived based on this 
interpretation accelerate the training speed. Further, the proposed method is able 
to recover the objective functions used by previous works under appropriate assumptions.
Also, by appropriately choosing the input representation for symbols, we show that
deep learning based systems can recover Gray coding through the training process.
The main differences in the proposed method when compared to existing AE based models 
are given in Table \ref{tab:characteristics}. In summary, this work introduces a 
deep learning based method for end to end design of communication systems which 
can systematically handle noise corruption of transmit symbols. The results show 
that the proposed method can produce consistently better models fast when compared 
to previous works.

\subsection{Notations}
Bold face lower-case letters (eg. $\textbf{x}$) denote column vector.
Bold face upper-case letters (eq. $\textbf{X}$) denote matrix.
Script face letters (eg. $\mathcal{S}$) denotes a set, $|\mathcal{S}|$
denotes the cardinality of the set $\mathcal{S}$. $f(\textbf{x};\bm{\theta})$
represents a function which takes in a vector $\textbf{x}$ and has parameters
$\bm{\theta}$. $\mathcal{D}_{KL}(p(X)||q(Y))$ denotes KL divergence between
random variables $X$ and $Y$ with distributions $p(\cdot)$ and $q(\cdot)$
respectively. $p_\theta(\cdot)$ represents a distribution with parameters $\theta$.
$\mathbb{E}_p$ is the expectation operator with respect to distribution $p$.
$\bm{0}_m$ represents an all zero vector of length $m$, $\bm{I}_m$ represents
identity matrix of dimension $m \times m$. $\mathcal{N}(\bm{\mu},\bm{\Sigma})$
is a multivariate Gaussian with mean vector $\bm{\mu}$ and covariance matrix $\bm{\Sigma}$.
The trace of a matrix $\textbf{A}$ is denoted by $tr(\textbf{A})$.

	\section{End to End Modeling of Communication Systems}  \label{sec:deep_comm_model}
A communication system can be seen as a model that recreates a copy of the message
which is sent by the transmitter at the receiver end. Let $\textbf{x} \in \mathcal{X}$ 
be the information
to be sent from the transmitter. Modern communication systems convert the data $\textbf{x}$
to a representation $\textbf{z} \in \mathcal{Z}$ which is suitable for transmission 
over a noisy channel. A corrupted version of $\textbf{z}$, denoted by $\hat{\textbf{z}}$
is received at the destination. The receiver tries to recover the best possible 
reconstruction of $\textbf{x}$ from the observed $\hat{\textbf{z}}$.

The transmitter can be viewed as a function which takes in the information $\textbf{x}$ 
and computes the intermediate representation $\textbf{z}$ as $\textbf{z} = f(\textbf{x})$.
The channel which corrupts $\textbf{z}$ can be represented as $\hat{\textbf{z}} = h(\textbf{z})$.
Here $h(\cdot)$ is a stochastic function which when applied on $\textbf{z}$ 
gives output ${\hat{\textbf{z}}}$.
Finally the receiver can be characterized as another function which computes the
best possible reconstruction of $\textbf{x}$ from $\hat{\textbf{z}}$ as $\hat{\textbf{x}}
= g(\hat{\textbf{z}})$.

\begin{figure}[h]
    \centering
\tikzstyle{data_block} = [draw, fill=white, circle, minimum size=0.75cm]
\tikzstyle{comm_block} = [draw, fill=white, rectangle, rounded corners, minimum height=0.8cm, minimum width=3.25cm]
\tikzstyle{nn_block} = [draw, fill=white, rectangle, minimum height=0.8cm, minimum width=3.25cm]
\tikzstyle{ch_block} = [draw, fill=gray!20, rectangle, rounded corners, minimum height=0.8cm, minimum width=3.25cm]
\tikzset{dotted_block/.style={draw=black!20!white, line width=1pt, dash pattern=on 1pt off 4pt on 6pt off 4pt,
            inner sep=6mm, rectangle, rounded corners}}

\begin{tikzpicture}[auto,>=stealth',every text node part/.style={align=center},scale=0.72, every node/.style={transform shape}]
    \begin{scope}[auto,node distance=1.0cm]
        \node [data_block] (data_src) {\textbf{x}};
        \node [nn_block, rotate=90, below of=data_src, node distance=1.2cm] (nn_ip) {Input Layer};
        \draw [->] (data_src) -- (nn_ip);
        \node [nn_block, rotate=90, below of=nn_ip] (nn_enc) {Encoder Layers};
        \draw [->] (nn_ip) -- (nn_enc);
        \node [nn_block, rotate=90, below of=nn_enc] (nn_norm) {Transmit Layer};
        \draw [->] (nn_enc) -- (nn_norm);
        \node [ch_block, left of=nn_norm, xshift=4cm, minimum width=4cm, minimum height=2.50cm] (channel) {Hardware Imperfections\\ \& \\ Physical Channel};
        \draw [->] (nn_norm) -- node{\textbf{z}} (channel);
        \node [nn_block, right of=channel, xshift= 2cm, rotate=90] (nn_decoders) {Decoder Layers};
        \draw [->] (channel) -- node{$\hat{\textbf{z}}$} (nn_decoders);
        \node [nn_block, below of=nn_decoders, rotate=90, yshift=-1cm,xshift=1cm] (nn_softmax) {Output Layer};
        \draw [->] (nn_decoders) -- (nn_softmax);
        \node [data_block, right of=nn_softmax, node distance=1.2cm] (data_sink) {$\hat{\textbf{x}}$};
        \draw [->] (nn_softmax) -- (data_sink);
        
        \node (tx_block) [dotted_block, fit = (nn_ip) (nn_enc) (nn_norm)] {};
        \node at (tx_block.north) [above, inner sep=3mm] {Transmitter Layers \\ $f(\textbf{x};\bm{\theta}_{T})$};
        \node (ch_impairment) [dotted_block, fit = (channel), minimum height = 4.0cm] {};
        \node at (ch_impairment.north) [above, inner sep = 3mm] {Channel Impairments \\ $h(\textbf{z};\bm{\theta}_C)$};
        \node (rx_block) [dotted_block, fit = (nn_decoders) (nn_softmax)] {};
        \node at (rx_block.north) [above, inner sep=3mm] {Receiver Layers  \\ $g(\hat{\textbf{z}},\bm{\theta}_{R})$};
        
    \end{scope}
\end{tikzpicture}
    \caption{Autoencoder based Communication System}
    \label{fig:deep_comm_sys}
\end{figure}
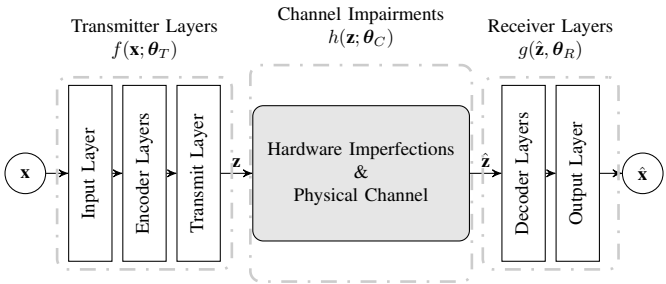

Following \cite{o2017introduction}, we can model a communication system as an
autoencoder.
The transmitter function is represented using a neural network parameterized by 
$\bm{\theta}_T$ such that $\textbf{z} = f(\textbf{x};\bm{\theta}_T)$ and the receiver 
function is represented using another neural network
parameterized by $\bm{\theta}_R$ such that $\hat{\textbf{x}} = g(\hat{\textbf{z}};
\bm{\theta}_R)$. However, the channel function $h(\cdot;\bm{\theta}_C)$, is typically
unknown in a communication system and is generally considered as a stochastic mapping 
from $\textbf{z}$ to $\hat{\textbf{z}}$. This channel function models both the hardware 
imperfections in the system as well as the channel impairments. Hence the communication
system can be represented as
\begin{align}
    \textbf{z} &= f(\textbf{x};\bm{\theta}_T) \\
    \hat{\textbf{z}} &= h(\textbf{z};\bm{\theta}_C) \\
    \hat{\textbf{x}} &= g(\hat{\textbf{z}};\bm{\theta}_R)
\end{align}
A schematic representation of the mentioned
design using neural network function approximators is provided in Fig. \ref{fig:deep_comm_sys}.

Let $\textbf{X} = \{ \textbf{x}_i\}_{i=1}^{N}$ represent the collection of input
symbols and $\hat{\textbf{X}} = \{ \hat{\textbf{x}}_i\}_{i=1}^{N}$ represent
the collection of decoded symbols.
The goal of an end to end communication system design is to find the parameters 
$\bm{\theta}_T$ and $\bm{\theta}_R$ such that 
\begin{align}
    \bm{\theta}_T, \bm{\theta}_R 
        &= \underset{\bm{\theta}_T,\bm{\theta}_R}{\arg \max} \: 
            \mathcal{G}(\textbf{X},\hat{\textbf{X}})    
            \label{eqn:train_obj_gain}
\end{align}
where $\mathcal{G}(\textbf{X},\hat{\textbf{X}})$ is a gain function which calculates 
how well the system is able to reconstruct the message in dataset $\textbf{X}$. 
Note that channel parameter $\bm{\theta}_C$ is not a learnable parameter and
hence not the part of the optimization objective
as it is dictated by channel.
Previous works 
\cite{o2017introduction,dorner2018deep,aoudia2018model,raj2018backpropagating} used one-hot
encoding to represent the message symbols $\textbf{x}$ and the gain is calculated
based on categorical cross-entropy over all the training samples. That is, $\mathcal{G}(\textbf{X},
\hat{\textbf{X}}) = \sum \limits_{\textbf{x}\in\textbf{X}} \log (p_\textbf{x})$, where
$p_\textbf{x} = p_\textbf{x}(\hat{\textbf{x}})$ corresponds to the normalized (to $1$) 
score given to the message $\textbf{x}$ from the output softmax layer.

In the following section, we discuss how to capture the latent code corruption
by channel into the model using the principles of variational inference and use 
the developments in the generative modeling capabilities of auto-encoder networks 
for simultaneously training the transmitter and the receiver.

	\section{Variational Inference perspective}
Efficient reconstruction $\hat{\textbf{x}}$ of message $\textbf{x}$ from the received 
representation $\hat{\textbf{z}}$ at receiver can be achieved if full knowledge of 
channel is available. However, the stochastic nature of channel function  $h(\cdot)$ 
and the 
lack of knowledge of the channel parameters $\bm{\theta_C}$ makes this goal challenging.
The joint density of the data that is transmitted $\textbf{x}$ and the received
signal $\hat{\textbf{z}}$ can be represented as
\begin{align}
    p(\textbf{x},\hat{\textbf{z}}) 
        = p(\textbf{x}) p(\hat{\textbf{z}}|\textbf{x})
        = p(\textbf{x}) p_{\bm{\theta}_C}(\hat{\textbf{z}}|\textbf{z}),
\end{align}
where we assume that transmitter provides a deterministic mapping from $\textbf{x}$
to $\textbf{z}$.
However, in an unknown channel scenario, the conditional density 
$p_{\bm{\theta}_C}(\hat{\textbf{z}}|\textbf{z})$, and in turn $p(\hat{\textbf{z}}|\textbf{x})$, is unknown.

\subsection{Graphical model for communication problem}
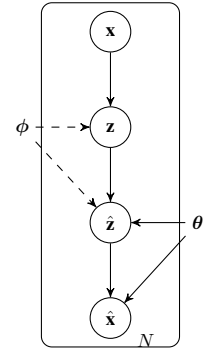
\begin{figure}[h]
        \centering
        \tikzstyle{variable_node} = [draw, fill=white, circle, minimum size=0.75cm]
\tikzset{plate/.style={draw=black, inner sep=10mm, rectangle, rounded corners}}

\begin{tikzpicture}[auto,>=stealth',every text node part/.style={align=center},scale=0.72, every node/.style={transform shape}]
    \begin{scope}[auto,node distance=1.0cm]
        \node [variable_node] (sym_x) {${\textbf{x}}$};
        \node [variable_node, below=1cm of sym_x] (var_z) {${\textbf{z}}$};
        \node [variable_node, below=1cm of var_z] (var_zhat) {$\hat{\textbf{z}}$};
        \node [variable_node, below=1cm of var_zhat] (var_x) {$\hat{\textbf{x}}$};
        \draw [->] (sym_x) -- (var_z);
        \draw [->] (var_z) -- (var_zhat);
        \draw [->] (var_zhat) -- (var_x);
        
        \node [right=1cm of var_zhat] (theta) {$\bm{\theta}$};
        \draw [->] (theta) -- (var_zhat);
        \draw [->] (theta) -- (var_x);

        \node [left=1cm of var_z] (phi) {$\bm{\phi}$};
        \draw [->,dashed] (phi) -- (var_z);
        \draw [->,dashed] (phi) -- (var_zhat);
        
        \node (base_plate) [plate, fit = (sym_x) (var_x)] {};

        \node [below right=-0.1cm and 0.1cm of var_x] (N) {$N$};        
    \end{scope}
\end{tikzpicture}
        \caption{Graphical model of proposed system}
        \label{fig:comm_vi_model}
    \end{figure}

    The problem of reliable communication can be cast into a graphical model
    as shown in Fig. \ref{fig:comm_vi_model}. Here, $\textbf{x}$ is the data
    and $\textbf{z}$ is the corresponding representation to be transmitted over the
    channel. 
    Here, we follow the standard plate notation of graphical models; variables $(\textbf{x},
    \hat{\textbf{x}}, \textbf{z}, \hat{\textbf{z}})$ are repeated $N$ times, while the parameters
    $(\bm{\phi, \theta})$ takes only a single realization in the problem.
    We use $\bm{\phi} = \{\bm{\theta}_T, \bm{\theta}_C\}$ to represent the parameters
    of the encoding process. In the graphical model, this is represented as $\textbf{z}$
    being influenced by $\textbf{x}$ and $\bm{\phi}$. The decoder with parameters $\bm{\theta}
    = \bm{\theta}_R$ acts on received representation $\hat{\textbf{z}}$ and produces a
    reconstruction of data.

    The main aim of a communication system is to identify the stochastic mapping of channel, from $\textbf{z}$
    to $\hat{\textbf{z}}$, and develop methods to retrieve the data $\textbf{x}$. In practical
    systems, it is often the case that the stochastic mapping of  channel is unknown
    and the distribution is difficult to compute.

    Variational Inference (VI) is a method from statistical learning for approximating difficult
    to compute probability densities \cite{blei2017variational}. VI deals with finding
    the conditional distribution of latent variables $\hat{\textbf{z}}$ given 
    $\textbf{x}$. Considering the joint density $p(\textbf{x},\hat{\textbf{z}}) = p(\textbf{x})
    p(\hat{\textbf{z}}|\textbf{x})$, 
    \textit{inference} in a Bayesian model amounts to conditioning on data and computing the
    posterior $p(\hat{\textbf{z}}|\textbf{x})$. Variational Inference applies optimization
    techniques to approximate this conditional density.
    
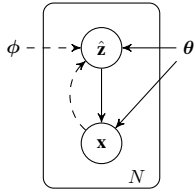
\begin{figure}[h]
        \centering
        \tikzstyle{variable_node} = [draw, fill=white, circle, minimum size=0.75cm]
\tikzset{plate/.style={draw=black, inner sep=8mm, rectangle, rounded corners}}

\begin{tikzpicture}[auto,>=stealth',every text node part/.style={align=center},scale=0.72, every node/.style={transform shape}]
    \begin{scope}[auto,node distance=1.0cm]
        \node [variable_node] (var_z) {$\hat{\textbf{z}}$};
        \node [variable_node, below=1cm of var_z] (var_x) {$\textbf{x}$};
        \draw [->] (var_z) -- (var_x);
        
        \node [right=1cm of var_z] (theta) {$\bm{\theta}$};
        \draw [->] (theta) -- (var_z);
        \draw [->] (theta) -- (var_x);

        \node [left=1cm of var_z] (phi) {$\bm{\phi}$};
        \draw [->,dashed] (phi) -- (var_z);
        \draw (var_x) edge[out=140,in=220,->, dashed] (var_z);

        \node (base_plate) [plate, fit = (var_z) (var_x)] {};

        \node [below right=0.1cm and 0.1cm of var_x] (N) {$N$};        
    \end{scope}
\end{tikzpicture}
        \caption{Graphical model of relationship between variables}
        \label{fig:latent_model}
    \end{figure}

    Recent developments in deep learning proposed the 
    use of variational inference for generative modeling. 
    Generative modeling refers to the process of producing valid samples from $p(\textbf{x})$.
    Consider the graphical model given in Fig. \ref{fig:latent_model}. Here, samples of
    $\textbf{x}$ are generated from a latent variable $\hat{\textbf{z}}$ and associated\
    parameters represented by $\bm{\theta}$. The solid lines
    denote the generative model $p_{\bm{\theta}}(\hat{\textbf{z}})p_{\bm{\theta}}(\textbf{x}|\hat{\textbf{z}})$.
    To generate valid samples of $\textbf{x}$, we first sample $\hat{\textbf{z}}$ and then
    use $\hat{\textbf{z}}$ and $\bm{\theta}$ to generate $\textbf{x}$. The dashed lines
    represent the inference procedure with variational approximation
    of the intractable posterior $p_{\bm{\theta}}(\hat{\textbf{z}}|\textbf{x})$.

    In \cite{kingma2013auto}, a stochastic
    optimization based method is proposed applying deep learning to first approximate 
    the inference $p(\hat{\textbf{z}}|\textbf{x})$ with appropriate prior on $p(\hat{\textbf{z}})$ 
    using an encoder network  $q_\phi(\hat{\textbf{z}}|\textbf{x})$. Then, a decoder 
    network $p_\theta(\textbf{x}|\hat{\textbf{z}})$ is used to compute the reconstruction
    $\hat{\textbf{x}}$ of message $\textbf{x}$ from $\hat{\textbf{z}}$. Here $\bm{\phi}$ and $\bm{\theta}$ 
    are the neural network parameters that will be learned during the training phase. 
    Given high capacity model (ie., neural networks with sufficient learning capability), 
    and good prior distribution $p(\hat{\textbf{z}})$, the model will approximate 
    the posterior ie., $q_\phi(\hat{\textbf{z}}|\textbf{x}) \approx p_{\bm{\theta}}(\hat{\textbf{z}}|
    \textbf{x})$. Because of the encoder-decoder structure present in the model, this
    method is generally known as \textit{Auto-encoding Variational Bayes (AVB)}.
    
    The expected marginal likelihood $p_{\theta}(\textbf{x})$ of datapoint $\textbf{x} \in 
    \mathcal{X}$, under an encoding function, $q_\phi(\cdot)$, can be computed as
    \begin{align}
        \mathbb{E}_{p(\textbf{x})} \log p_\theta(\textbf{x}) 
                        = \mathcal{L}_{\theta,\phi}(\textbf{x}) + 
                            \mathbb{E}_{p(\textbf{x})} 
                                \mathcal{D}_{KL}(q_\phi(\hat{\textbf{z}}|\textbf{x})||
                                            p_\theta(\hat{\textbf{z}}|\textbf{x})),
                                            \label{eqn:loglikelihood}
    \end{align}
    where
    \begin{align}
        \mathcal{L}_{\theta,\phi}(\textbf{x}) 
            &= \mathbb{E}_{p(\textbf{x})} 
                    \mathbb{E}_{q_\phi(\hat{\textbf{z}}|\textbf{x})} 
                    \left(
                        \log \frac{p_\theta(\textbf{x},\hat{\textbf{z}})}
                                {q_\phi(\hat{\textbf{z}}|\textbf{x})}
                    \right) \label{eqn:elbo_def}
    \end{align}
    is commonly referred as \textit{Evidence Lower Bound} (ELBO) or \textit{Variational
    Lower Bound} and
    \begin{align}
        \mathcal{D}_{KL}(
            q_\phi(\hat{\textbf{z}}|\textbf{x})||
            p_\theta(\hat{\textbf{z}}|\textbf{x}) )
        &= \mathbb{E}_{q_\phi(\hat{\textbf{z}}|\textbf{x})} \left(
                                \log \frac{q_\phi(\hat{\textbf{z}}|\textbf{x})}
                                        {p_\theta(\hat{\textbf{z}}|\textbf{x})}
                            \right)
    \end{align}
    is the KL-divergence between the approximating and actual  distributions.
    Please see Appendix \ref{app:loglikehood} for details on (\ref{eqn:loglikelihood}).
    By
    re-arranging (\ref{eqn:loglikelihood}) and noting that $D_{KL}(Y_1||Y_2) \geq 0$
    for any two random variables $\{Y_1,Y_2\}$, we can see that 
    $\log p_\theta(\textbf{x}) \geq \mathcal{L}_{\theta,\phi}(\textbf{x})$. Therefore,
    the likelihood of reconstruction $\log p_\theta(\textbf{x})$ is lower bounded by
    (\ref{eqn:elbo_def}) (hence the name \textit{ELBO}). Since
    it is difficult to compute the value of ${p_\theta(\hat{\textbf{z}}|\textbf{x})}$,
    Variational Inference
    tries to maximize this alternative quantity $\mathbb{E}_{p(\textbf{x})}  
    \log p_\theta(\textbf{x}) - \mathbb{E}_{p(\textbf{x})} \mathcal{D}_{KL}
    (q_\phi(\hat{\textbf{z}}|\textbf{\textbf{x}})||p_\theta(\hat{\textbf{z}}|\textbf{x}))$ 
    by maximizing the ELBO $\mathcal{L}_{\theta,\phi}(\textbf{x})$. 

    Following from (\ref{eqn:elbo_def}), the maximization objective ELBO 
    $\mathcal{L}_{\theta,\phi}(\textbf{x})$ can be re-arranged as
    \begin{align}
\ifCLASSOPTIONtwocolumn
        \mathcal{L}_{\theta,\phi}&(\textbf{x}) 
            \nonumber \\
\else
        \mathcal{L}_{\theta,\phi}(\textbf{x}) 
\fi
            &= \mathbb{E}_{p(\textbf{x})} \mathbb{E}_{q_\phi(\hat{\textbf{z}}|\textbf{x})} 
                        \log p_\theta(\textbf{x},\hat{\textbf{z}}) -
                \mathbb{E}_{p(\textbf{x})} \mathbb{E}_{q_\phi(\hat{\textbf{z}}|\textbf{x})} 
                        \log {q_\phi(\hat{\textbf{z}}|\textbf{x})} \nonumber \\
            &= \mathbb{E}_{p(\textbf{x})} \mathbb{E}_{q_\phi(\hat{\textbf{z}}|\textbf{x})} 
                        \log \left( p_\theta(\textbf{x}|\hat{\textbf{z}}) p(\hat{\textbf{z}}) \right) -
                \mathbb{E}_{p(\textbf{x})} \mathbb{E}_{q_\phi(\hat{\textbf{z}}|\textbf{x})} 
                        \log {q_\phi(\hat{\textbf{z}}|\textbf{x})} \nonumber \\
            &= \mathbb{E}_{p(\textbf{x})} \mathbb{E}_{q_\phi(\hat{\textbf{z}}|\textbf{x})} 
                        \log p_\theta(\textbf{x}|\hat{\textbf{z}}) -
                \mathbb{E}_{p(\textbf{x})} \mathbb{E}_{q_\phi(\hat{\textbf{z}}|\textbf{x})} \left(  
                        \log\frac{ {q_\phi(\hat{\textbf{z}}|\textbf{x})}}
                            {p(\hat{\textbf{z}})}  \right) \nonumber \\
            &= \underbrace{\mathbb{E}_{p(\textbf{x})} \mathbb{E}_{q_\phi(\hat{\textbf{z}}|\textbf{x})} 
                        \log p_\theta(\textbf{x}|\hat{\textbf{z}})}
                            _{\text{reconstruction likelihood}} -
                \underbrace{\mathbb{E}_{p(\textbf{x})} \mathcal{D}_{KL}(q_\phi(\hat{\textbf{z}}|\textbf{x})||
                        p(\hat{\textbf{z}}))}
                            _{\text{KL loss}}.    \label{eqn:elbo_split}
    \end{align}
    Hence, the objective of maximizing ELBO is equivalent to maximizing
    the penalized likelihood of reconstruction of $\textbf{x}$ from $\hat{\textbf{z}}$
    where is the penalty is the KL-divergence
    between the inference density approximation $q_\phi(\hat{\textbf{z}}|\textbf{x})$
    and assumed prior $p_{\theta}(\hat{\textbf{z}})$.

    From Fig. \ref{fig:deep_comm_sys}, $\bm{\theta}_T$ and $\bm{\theta}_R$ are the only 
    learnable parameters in this system and $\bm{\theta}_C$ represents the unknown parameters 
    of the channel along with the stochastic channel function $h(\cdot)$. From the 
    model presented in Fig. \ref{fig:comm_vi_model}, we have $\bm{\phi} 
    = \{\bm{\theta}_T, \bm{\theta}_C\}$ and $\bm{\theta} = \bm{\theta_R}$. 
    In AVB, the encoder network $\mathcal{E}_\phi$ is used to learn the parameters to 
    compute $\textbf{z}$ from given symbol $\textbf{x}$. Then, a stochastic channel
    function is applied on $\textbf{z}$ to sample $\hat{\textbf{z}}$ which is used 
    by the decoder network $\mathcal{D}_\theta$ to recreate $\textbf{x}$. 
    Hence,
    \begin{alignat}{3}
        \textbf{z} 
            &= f(\textbf{x};\bm{\theta}_T),
            &&= \mathcal{E}_\phi(\textbf{x}) \\
        q_\phi(\hat{\textbf{z}}|\textbf{x}) 
            &= h(\textbf{z};\bm{\theta_C}) \\
        p_\theta(\hat{\textbf{x}}|\hat{\textbf{z}}) 
            &= g( \hat{\textbf{z}};\bm{\theta}_R )
            &&= \mathcal{D}_\theta(\hat{\textbf{z}}).
    \end{alignat}

    The effect of the encoder $\mathcal{E}_\phi$ and the stochastic channel function 
    which together transform the message $\textbf{x}$ to a representation $\hat{\textbf{z}}$ 
    which suffered corruption from the channel is approximated by $q_{\phi}(\hat{\textbf{z}}
    |\textbf{x})$.
    The output of the decoder $\mathcal{D}_\theta$
    is a distribution over all the possible messages computed after
    observing $\hat{\textbf{z}}$ and is represented as $p_\theta(\hat{\textbf{x}}|
    \hat{\textbf{z}})$.

    Finally, the objective of the optimization problem (\ref{eqn:train_obj_gain}) to 
    train end to end communication system having the model discussed above can be
    written as
    \begin{align}
        \bm{\theta}_T, \bm{\theta}_R 
            &= \underset{\bm{\theta}_T,\bm{\theta}_R}{\arg \max} \: 
                    \mathcal{L}_{\bm{\theta},\bm{\phi}}(\textbf{x}),
                    \label{exp:obj_general}
    \end{align}
    over all $\textbf{x} \in \textbf{X}$, the set of available training points.

\subsection{Reconstruction likelihood}
    The first term in maximizing objective ELBO (\ref{eqn:elbo_split}) accounts for the 
    capability of the end to end system to successfully reproduce the intended message 
    $\textbf{x}$ at the receiver end. The exact expression for reconstruction likelihood 
    $\mathbb{E}_{p(\textbf{x})} \mathbb{E}_{q_\phi(\hat{\textbf{z}}|\textbf{x})} 
    \log p_\theta(\textbf{x}|\hat{\textbf{z}})$ 
    depends on how the message $\textbf{x}$ is represented in the system.

    Previous works on end to end design of communication systems \cite{o2017introduction,
    dorner2018deep,raj2018backpropagating,aoudia2018model} used one-hot encoding
    to represented each message $\textbf{x} \in \mathcal{X}$. With $|\mathcal{X}| = M$,
    one-hot encoding uses a vector of length $M$ with all entries $0$ except a $1$ 
    for the position corresponding to the message. The softmax output layer of the receiver 
    also produces a $M$ length vector, which sums to $1$. 
    If this representation of $\textbf{x}$ is used, the reconstruction term 
    in (\ref{eqn:elbo_split}) takes the form of negative categorical cross entropy and 
    can be written as
    \begin{align}
    \mathbb{E}_{p(\textbf{x})} \mathbb{E}_{q_\phi(\hat{\textbf{z}}|\textbf{x})} 
        \log p_\theta(\textbf{x}|\hat{\textbf{z}})
        &= \sum \limits_{\textbf{x} \in \textbf{X}}  \log(p_{\textbf{x}}),    \label{eqn:recon_onehot}
    \end{align}
    where $p_\textbf{x}$ corresponds to the normalized (to $1$) score given to the message 
    $\textbf{x}$ by the receiver $\mathcal{D}_\theta(\cdot)$'s softmax output layer.

    Another way of representing the message is to directly use the binary representation
    of the message. For $|\mathcal{X}| = M$, we need a block length of atleast $d = 
    \lceil \log_2M \rceil$ to represent (uncoded) message $\textbf{x}$. Under this 
    representation, $\textbf{x}$ is a vector of length $d$ with multiple entries of $0$s
    and $1$s. The output layer of decoder should also be properly modified to output
    the corresponding values. In this case, a popular choice for output layer activation 
    function is to use sigmoid activation, which assigns a value between $0$ and $1$ 
    for each of the entries in reconstruction. Hence $p_\theta(\textbf{x}|\hat{\textbf{z}})$ 
    becomes a multivariate Bernoulli distribution of length $b$ with element probabilities
    computed from $\hat{\textbf{z}}$. The reconstruction likelihood becomes negative of
    binary cross entropy as in \cite{kingma2013auto} and can be computed as,
    \begin{align}
\ifCLASSOPTIONtwocolumn
    \mathbb{E}_{p(\textbf{x})} &\mathbb{E}_{q_\phi(\hat{\textbf{z}}|\textbf{x})}
        \log p_\theta(\textbf{x}|\hat{\textbf{z}})  \nonumber \\
\else
    \mathbb{E}_{p(\textbf{x})} \mathbb{E}_{q_\phi(\hat{\textbf{z}}|\textbf{x})}
            \log p_\theta(\textbf{x}|\hat{\textbf{z}})
\fi
        &= \sum \limits_{\textbf{x} \in \textbf{X}} 
                \sum \limits_{i = 1}^{d} \log p_\theta(x_i|\hat{\textbf{z}})
        = \sum \limits_{\textbf{x} \in \textbf{X}} 
                \sum \limits_{i = 1}^{d} \log p(x_i;\hat{x}_i) \nonumber \\
        &= \sum \limits_{\textbf{x} \in \textbf{X}} \sum \limits_{i = 1}^{d}
                \left( 
                    x_i \log \hat{x}_i +
                    (1-x_i) \log (1-\hat{x}_i)
                \right).    \label{eqn:recon_bernoulli}
    \end{align}

    While one-hot representation with categorical cross entropy is a popular choice of
    loss function for classification tasks, the binary message representation with binary 
    cross entropy is scalable to a learn for a very large number of messages
    \footnote{While one-hot encoding requires $M$ nodes at the inputs layer, binary
    representation only requires only $\lceil \log_2 M\rceil$ nodes at inputs.}. One should select the 
    appropriate representation for messages while keeping these constraints in mind.
    In Sec \ref{sec:results}, we show that by using (\ref{eqn:recon_bernoulli}) instead of
    (\ref{eqn:recon_onehot}) on (\ref{eqn:elbo_split}), the models can
    be taught the concept of \textit{Gray Coding} without any other explicit criterion.

    Note that (\ref{eqn:elbo_split}) composes of two terms and in the succeeding subsections,
    we discuss the second term and its impact. Also, note that when the second 
    term in (\ref{eqn:elbo_split}) is a constant, the first term will be the optimization
    objective and we recover the results in \cite{o2017introduction,
    dorner2018deep,raj2018backpropagating,aoudia2018model}.

\subsection{KL-loss for AWGN channel}
    The Additive White Gaussian Noise (AWGN) channel is a widely used channel model
    to represent the corruption incurred to the transmitted signal
    in communication systems. For a $\textbf{z}$
    of dimensions $m$, Gaussian corruption with noise power $\sigma_n^2$ per component
    is modeled as
    \begin{align}
        \hat{\textbf{z}} = \textbf{z} + \bm{n},     \label{eqn:awgn_model}
    \end{align}
    where $\textbf{n} \sim \mathcal{N}(\bm{0}_m,\sigma_n^2 \bm{I}_m)$; $\bm{0}$ 
    is an all zero vector of dimension $m$ and $\bm{I}$ is an identity matrix of
    dimension $m \times m$. Taking a Gaussian prior of $p(\hat{\textbf{z}}) = \mathcal{N}(\bm{0}_m,
    \sigma_0^2 \bm{I}_m)$, the KL Loss in (\ref{eqn:elbo_split}) for AWGN channel
    can be computed as
    \begin{align}
        \mathcal{D}_{KL}(q_\phi(\hat{\textbf{z}}|\textbf{x})||p(\hat{\textbf{z}}))
            &= \frac{1}{2\sigma_0^2} \sum \limits_{j=1}^{m} z_j^2
                - \frac{m}{2} \left( 1 - \frac{\sigma_n^2}{\sigma_0^2} 
                + \log \frac{\sigma_n^2}{\sigma_0^2} \right). \label{eqn:klloss_awgn}
    \end{align}
    Please refer Appendix \ref{app:awgn_obj} for the derivation.
    Depending on the representation
    used for symbols in the model, (\ref{eqn:klloss_awgn}) can be combined with 
    (\ref{eqn:recon_onehot}) (in case of one-hot representation) or with (\ref{eqn:recon_bernoulli})
    (in case of binary representation) to get appropriate objective function for
    training the model in AWGN channel.

    Considering the case of one-hot encoding as in \cite{o2017introduction,dorner2018deep,
    raj2018backpropagating,aoudia2018model}, the ELBO objective to be maximized
    ie., (\ref{eqn:elbo_split}) can then be computed as
    \begin{align}
        \sum \limits_{\textbf{x} \in \textbf{X}} 
            \left( \log(p_{\textbf{x}}) 
            - \frac{1}{2\sigma_0^2} \sum \limits_{j=1}^{m} z_j^2
            + \frac{m}{2} \left( 1 - \frac{\sigma_n^2}{\sigma_0^2} 
                + \log \frac{\sigma_n^2}{\sigma_0^2} \right) \right)
    \end{align}
    As the noise power per component $\sigma_n^2$ and the prior variance $\sigma_0^2$
    are constant in the problem, the final objective to maximize can be written
    as 
    \begin{align}
        \underset{\bm{\theta}_T,\bm{\theta}_R}{\max} \left\{ 
            \sum \limits_{\textbf{x} \in \textbf{X}} \left(    
                \log(p_{\textbf{x}}) 
                - \frac{1}{2\sigma_0^2} \sum \limits_{j=1}^{m} z_j^2 
                \right)
            \right\}.
            \label{exp:obj_awgn_1hot}
    \end{align}

    The first term in the derived objective
    (\ref{exp:obj_awgn_1hot}) is negative of the categorical cross entropy. Previous
    works in \cite{o2017introduction, dorner2018deep,raj2018backpropagating,aoudia2018model}
    considered only this term for optimization at a constant training SNR\footnote{
    The SNR in this case is defined as $SNR = \frac{1}{m\sigma_n^2} \sum \limits_{j=1}^{m} z_j^2$.}.
    The second term connects the signal power $\sum \limits_{j=1}^{m} z_j^2$ and noise power 
    to the design. At a specified noise power $\sigma_n^2$ per component, maximization of 
    the above objective brings in the concept of using less power for signaling. 
    Hence, the derived objective optimizes the signaling such that 
    a tradeoff is achieved between minimizing the transmit power and maximizing 
    the reconstruction likelihood.
    If we assume a constant training $SNR$ scenario, the second term becomes 
    a constant and we recover the objective used in 
    \cite{o2017introduction, dorner2018deep,raj2018backpropagating,aoudia2018model}.

    Comparing the derived objective (\ref{exp:obj_awgn_1hot}) to the objective used in AE 
    based communication
    systems design popularized by \cite{o2017introduction} points to some interesting
    observations. The main difference between the proposed method and AE based design
    are that AE based designs use a normalization layer as the last layer in transmitter
    to control the power used for signaling. By choosing a particular SNR, $\gamma$,
    to train at, the objective of these models is to maximize the reconstruction
    likelihood alone. Let $\sigma_n^2$ be the noise power per component of the transmission
    from the channel and $m$ be the number of components. Then the objective to
    optimize, with power constraint from the normalization layer, becomes,
    \begin{align}
        \underset{\bm{\theta}_T,\bm{\theta}_R}{\max} \quad
            {\mathbb{E}_{p(\textbf{x})}\mathbb{E}_{q(\hat{\textbf{z}}|\textbf{x})}}
            \log p_{\bm{\theta}_R}(\textbf{x}|\hat{\textbf{z}}) \nonumber \\
            \text{sub. to} \qquad \mathbb{E}_{p(\textbf{x})}\textbf{z}^T\textbf{z}
                = m \sigma_n^2 \gamma.
    \end{align}

    Introducing Lagrangian multiplier, we can rewrite the above optimization objective as,
    \begin{align}
        \underset{\bm{\theta}_T,\bm{\theta}_R}{\max} \quad \left\{
            {\mathbb{E}_{p(\textbf{x})}\mathbb{E}_{q(\hat{\textbf{z}}|\textbf{x})}}
            \log p_{\bm{\theta}_R}(\textbf{x}|\hat{\textbf{z}}) 
            - \lambda_L \mathbb{E}_{p(\textbf{x})} \textbf{z}^T \textbf{z}
            - \lambda_L m \sigma_n^2 \gamma
            \right\},
    \end{align}
    where $\lambda_L$ is the Lagrangian multiplier.
    Removing the problem independent constants, this can be re-written as,
    \begin{align}
        \underset{\bm{\theta}_T,\bm{\theta}_R}{\max}  \quad \left\{
            \mathbb{E}_{p(\textbf{x})}
            \left( \mathbb{E}_{q(\hat{\textbf{z}}|\textbf{x})}
                \log p_{\bm{\theta}_R}(\textbf{x}|\hat{\textbf{z}}) 
                - \lambda_L \textbf{z}^T \textbf{z}
            \right)
            \right\}. \label{eqn:commae_lagrangian}
    \end{align}

    Comparing this with the objective derived in (\ref{exp:obj_awgn_1hot}), we can
    observe that AE based models \cite{o2017introduction} are also following a similar 
    objective function to maximize with $\lambda_L = \frac{1}{2\sigma_0^2}$. 
    In other words, while the works in
    \cite{o2017introduction, dorner2018deep,raj2018backpropagating,aoudia2018model}
    impose hard constraints, this
    work imposes a soft constraint. 
    
    Recent developments in research to incorporate hard constraints to deep learning
    problems suggest that imposing a hard constraint on a deep learning problem 
    may not lead to desired performance.
    The work in \cite{kervadec2019constrained} suggests that hard constraints
    should mostly be avoided and instead proposes to use differentiable penalties
    in loss functions, similar to our approach.
    In \cite{marquez2017imposing}, authors show that even though hard constraints
    bring in nice theoretical benefits, the resulting method can end up being computationally
    complex (like the addition of normalization layer at the output of the encoder as done
    in existing works \cite{o2017introduction, dorner2018deep,raj2018backpropagating,aoudia2018model}).
    Further, the promised benefits may not manifest in practical problems.
    Later, in the Results section, we show that the proposed method of soft constraints
    on loss function yields faster training that the hard constraint approach adopted
    in previous works.
    
\subsection{KL-loss for Rayleigh Block Fading (RBF) channel}
    One of the most widely used model to capture the fading effects during
    signal transmission is Rayleigh Block Fading.
    Under Rayleigh Block Fading (RBF) model, the corrupted signal $\hat{\textbf{z}}$
    can be modeled as
    \begin{align}
        \hat{\textbf{z}} = h \textbf{z} + \bm{n},     \label{eqn:rbf_model}
    \end{align}
    where $h \sim \mathcal{CN}(0,1)$ and $\textbf{n} \sim \mathcal{N}(\bm{0}_m,
    \sigma_n^2 \bm{I}_m)$ or equivalently \cite{aoudia2018model}
    \begin{align}
        \hat{\textbf{z}} \sim \mathcal{N}\left( \bm{0}, 
                \frac{1}{2}\left( \textbf{z}\textbf{z}^T 
                    - \textbf{Jz}\textbf{z}^T\textbf{J} \right) +
                    \sigma_n^2 \bm{I}_m \right),  \label{eqn:rbf_cov_model}
    \end{align}
    where $\textbf{J}$ is the matrix defined by $\textbf{J} = \begin{bmatrix}
    \bm{0}_{m/2} & -\bm{I}_{m/2} \\ \bm{I}_{m/2} & \bm{0}_{m/2} \end{bmatrix}$ with 
    $\bm{0}_{m/2}$ is square zero matrix of dimension $m/2$ and $\bm{I}_{m/2}$ 
    identity matrix of dimension $m/2$ 
    \footnote{Note that while implementing in DNN, we split complex $\textbf{z}$ 
    in to real and imaginary components and stack them into a column vector of dimension
    $m$. Hence $m$ is always even in the model.}.
    If the only knowledge we have about the channel is that it can
    be well modeled by a distribution with finite variance, then the prior choice
    should reflect this information. In this context, a normal prior is the maximum
    entropy prior.
    Hence, taking a prior of $p(\hat{\textbf{z}}) = \mathcal{N}(\bm{0}_m,
    \sigma_0^2 \bm{I}_m)$,
    the KL Loss in (\ref{eqn:elbo_split}) for this case can be computed as
\ifCLASSOPTIONtwocolumn
    \begin{align}
        \mathcal{D}_{KL}(q_\phi(\hat{\textbf{z}}|\textbf{x})||p(\hat{\textbf{z}}))
            &= \frac{1}{2\sigma_0^2} \sum\limits_{j=1}^{m} z_j^2 
            - \frac{m}{2} \left( 1 - \frac{\sigma_n^2}{\sigma_0^2} 
                + \log \frac{\sigma_n^2}{\sigma_0^2}\right)     \nonumber \\
            & \qquad 
            - \log \left( 1 + \frac{1}{2\sigma_n^2} \sum \limits_{j=1}^{m} z_j^2 \right).
                \label{eqn:klloss_rbf}
    \end{align}
\else
    \begin{align}
        \mathcal{D}_{KL}(q_\phi(\hat{\textbf{z}}|\textbf{x})||p(\hat{\textbf{z}}))
            &= \frac{1}{2\sigma_0^2} \sum\limits_{j=1}^{m} z_j^2 
            - \frac{m}{2} \left( 1 - \frac{\sigma_n^2}{\sigma_0^2} 
                + \log \frac{\sigma_n^2}{\sigma_0^2}\right)
            - \log \left( 1 + \frac{1}{2\sigma_n^2} \sum \limits_{j=1}^{m} z_j^2 \right).
                \label{eqn:klloss_rbf}
    \end{align}
\fi
    Please refer to Appendix \ref{app:rbf_obj} for the detailed derivation.
    Depending on the representation
    used for symbols in the model, (\ref{eqn:klloss_rbf}) can be combined with 
    (\ref{eqn:recon_onehot}) (in case of one-hot representation) or with (\ref{eqn:recon_bernoulli})
    (in case of binary representation) to get appropriate objective function for
    training the model in RBF channel.

    Considering one-hot encoding
    and removing the constant terms in the problem, the final ELBO objective (\ref{eqn:elbo_split})
    to maximize for training an end to end communication system in an RBF channel
    can be written as
\ifCLASSOPTIONtwocolumn
    \begin{align}
        \underset{\bm{\theta}_T,\bm{\theta}_R}{\max} &\left\{ 
            \sum \limits_{\textbf{x} \in \textbf{X}}
            \left( \log(p_{\textbf{x}}) 
            - \frac{1}{2\sigma_0^2} \sum \limits_{j=1}^{m} z_j^2  
            \right. \right. \nonumber \\ &\qquad \qquad \qquad \left. \left.
            + \log \left( 1 + \frac{1}{2\sigma_n^2} \sum \limits_{j=1}^{m} z_j^2 \right)
            \right) \right\}. \label{exp:obj_rbf_1hot}
    \end{align}
\else
    \begin{align}
        \underset{\bm{\theta}_T,\bm{\theta}_R}{\max} &\left\{ 
            \sum \limits_{\textbf{x} \in \textbf{X}}
            \left( \log(p_{\textbf{x}}) 
            - \frac{1}{2\sigma_0^2} \sum \limits_{j=1}^{m} z_j^2  
            + \log \left( 1 + \frac{1}{2\sigma_n^2} \sum \limits_{j=1}^{m} z_j^2 \right)
            \right) \right\}. \label{exp:obj_rbf_1hot}
    \end{align}
\fi
    This objective is
    slightly different from the AWGN objective (\ref{exp:obj_awgn_1hot})
    due to an additional term similar to capacity.
    Similar to the case of AWGN channel objective, we can see that at constant 
    SNR condition, we recover the 
    objective function used in \cite{o2017introduction, dorner2018deep,raj2018backpropagating,
    aoudia2018model}. Interestingly, in the special case of $m=2$, the new term in this objective (the
    third term in (\ref{exp:obj_rbf_1hot})) is equivalent to the AWGN channel capacity.
    Maximizing this objective optimizes the system to improve the channel capacity (third
    term) while minimizing the signaling energy (second term) and at the same time 
    improving reconstruction loss (first term). This intuitively fits with the 
    objective of communication systems - maximize the capacity while using minimum
    signaling power.

\subsection{Constellation learning and mutual information}
    Mutual information between the data to send $\textbf{x}$ and the received
    symbol $\hat{\textbf{z}}$ can be written as,
    \begin{align*}
        I(\textbf{X};\hat{\textbf{Z}})
            &= \mathbb{E}_{q_{\phi}(\textbf{x},\hat{\textbf{z}})}
                    \log \frac{q_{\phi}(\hat{\textbf{z}}|\textbf{x})}{q_{\phi}(\hat{\textbf{z}})}
    \end{align*}
    Following this definition, a lower bound on the mutual information can be obtained as
    \begin{align}
        I(\textbf{X};\hat{\textbf{Z}})
            &= \mathbb{E}_{q_{\phi}(\textbf{x},\hat{\textbf{z}})}
                    \log \frac{q_{\phi}(\textbf{x}|\hat{\textbf{z}})}{p(\textbf{x})} \nonumber \\
\ifCLASSOPTIONtwocolumn
            &= \mathbb{E}_{p(\textbf{x})} \mathbb{E}_{q_\phi(\hat{\textbf{z}}|\textbf{x})}
                    \log \frac{p_\theta(\textbf{x}|\hat{\textbf{z}})}{p(\textbf{x})} 
                \nonumber \\ & \qquad \qquad
                    + \mathbb{E}_{q_\phi(\hat{\textbf{z}})} \mathbb{E}_{q_\phi(\textbf{x}|\hat{\textbf{z}})}
                        \log \frac{q_\phi(\textbf{x}|\hat{\textbf{z}})}
                            {p_\theta(\textbf{x}|\hat{\textbf{z}})} \nonumber \\
            &= \mathbb{E}_{p(\textbf{x})} \mathbb{E}_{q_\phi(\hat{\textbf{z}}|\textbf{x})}
                    \log p_\theta(\textbf{x}|\hat{\textbf{z}})
                        + \mathbb{E}_{p(\textbf{x})} \log \frac{1}{p(\textbf{x})} 
                    \nonumber \\ &\qquad \qquad  
                        + \mathbb{E}_{q_\phi(\hat{\textbf{z}})}
                            \mathcal{D}_{KL}\left(
                                q_\phi(\textbf{x}|\hat{\textbf{z}})||p_\theta(\textbf{x}|\hat{\textbf{z}})
                            \right) \nonumber \\
\else
            &= \mathbb{E}_{p(\textbf{x})} \mathbb{E}_{q_\phi(\hat{\textbf{z}}|\textbf{x})}
            \log \frac{p_\theta(\textbf{x}|\hat{\textbf{z}})}{p(\textbf{x})} 
            + \mathbb{E}_{q_\phi(\hat{\textbf{z}})} \mathbb{E}_{q_\phi(\textbf{x}|\hat{\textbf{z}})}
                \log \frac{q_\phi(\textbf{x}|\hat{\textbf{z}})}
                    {p_\theta(\textbf{x}|\hat{\textbf{z}})} \nonumber \\
            &= \mathbb{E}_{p(\textbf{x})} \mathbb{E}_{q_\phi(\hat{\textbf{z}}|\textbf{x})}
            \log p_\theta(\textbf{x}|\hat{\textbf{z}})
                + \mathbb{E}_{p(\textbf{x})} \log \frac{1}{p(\textbf{x})} 
                + \mathbb{E}_{q_\phi(\hat{\textbf{z}})}
                    \mathcal{D}_{KL}\left(
                        q_\phi(\textbf{x}|\hat{\textbf{z}})||p_\theta(\textbf{x}|\hat{\textbf{z}})
                    \right) \nonumber \\
\fi
            &\geq \mathbb{E}_{p(\textbf{x})} \mathbb{E}_{q_\phi(\hat{\textbf{z}}|\textbf{x})}
                    \log p_\theta(\textbf{x}|\hat{\textbf{z}}).
                    \label{eqn:mi_lower}
    \end{align}
    Similarly, an upper bound on mutual information can be obtained as,
    \begin{align}
        I(\textbf{X};\hat{\textbf{Z}})
            &= \mathbb{E}_{q_{\phi}(\textbf{x},\hat{\textbf{z}})}
                    \log \frac{q_{\phi}(\hat{\textbf{z}}|\textbf{x})}{q_{\phi}(\hat{\textbf{z}})} \nonumber \\
            &= \mathbb{E}_{q_{\phi}(\textbf{x},\hat{\textbf{z}})} 
                    \log \frac{q_\phi(\hat{\textbf{z}}|\textbf{x})}{p(\hat{\textbf{z}})}
                    + \mathbb{E}_{q_{\phi}(\textbf{x},\hat{\textbf{z}})}
                        \log \frac{p(\hat{\textbf{z}})}{q_\phi(\hat{\textbf{z}})} \nonumber \\
            &= \mathbb{E}_{p(\textbf{x})} \mathbb{E}_{q_{\phi}(\hat{\textbf{z}}|\textbf{x})} 
                    \log \frac{q_\phi(\hat{\textbf{z}}|\textbf{x})}{p(\hat{\textbf{z}})}
                    - \mathbb{E}_{q_{\phi}(\hat{\textbf{z}})} 
                        \log \frac{q_\phi(\hat{\textbf{z}})}{p(\hat{\textbf{z}})} \nonumber \\
            &= \mathbb{E}_{p(\textbf{x})} 
                    \mathcal{D}_{KL}(q_\phi(\hat{\textbf{z}}|\textbf{x})||p(\hat{\textbf{z}}))
                - \mathcal{D}_{KL}(q_\phi(\hat{\textbf{z}})||p(\hat{\textbf{z}})) \nonumber \\
            & \leq \mathbb{E}_{p(\textbf{x})} 
                    \mathcal{D}_{KL}(q_\phi(\hat{\textbf{z}}|\textbf{x})||p(\hat{\textbf{z}})).
                    \label{eqn:mi_upper}
    \end{align}
    Comparing the ELBO objective used to train the proposed system (\ref{eqn:elbo_split}) and the
    bounds derived in (\ref{eqn:mi_lower}) and (\ref{eqn:mi_upper}), we can observe that
    the objective (\ref{eqn:elbo_split}) simultaneously tries to maximize a lower
    bound of mutual information and minimize an upper bound of the same. The weight given to
    the objective of minimizing the upper bound can be controlled by the parameter $\sigma_0^2$
    and hence, in the training process, more importance can be given to maximizing the lower bound.
    As discussed previously, the AE-based end-to-end communication systems also have similar
    objective function as ours and hence they also follow similar procedure of maximizing
    lower bound on mutual information and minimizing a weighted upper bound.
    Note, it is the upper bound
    minimizing term which brings in the concept of power control to the model.
    By controlling the mutual information between $\hat{\textbf{z}}$ and $\textbf{x}$,
    the models avoid trivially scaling the transmit symbols to avoid the effect
    of channel distortion. However, AE based methods \cite{o2017introduction,dorner2018deep,
    aoudia2018model} having hard constraint place more emphasis on minimizing the upper 
    bound on mutual information.

\subsection{Discussion}
    In this section, we presented an approach for end to end designing of communication
    systems based on the principles of variational inference and the recent developments
    in generative modeling with deep neural networks. 
    We showed how any prior information about the channel, either in the form of channel 
    parameters or in the functional form of the channel,
    can be appropriately incorporated for designing the objective function for optimization
    through (\ref{eqn:elbo_split}). 
    We also provided two examples, with the case of AWGN and RBF channel models.
    Previous works had to include an additional normalization
    layer at the transmitter output to control the power of the transmit symbols, which
    otherwise can become very high. 
    This is because, the objective functions used by the learning agents in those works
    have no incentive for controlling the transmit power.
    However, our proposed method yields objective
    functions which implicitly take care of transmit power control and hence eliminate
    the need for an additional normalization layer at the transmitter output.
        
    Generalizing beyond AWGN and RBF channel models, the method we proposed in this section
    can be
    applied to additive non-Gaussian noise channels as well as  other generalized fading 
    channel models using suitable prior. In the scenarios where such additional
    knowledge of the channel is available, the KL-loss in (\ref{eqn:elbo_split}) has
    to be computed with appropriate prior to obtain the objective function for training.
    As an example for other noise scenarios, we derive the loss function 
    for Additive Independent Laplace Noise (AILN) environments in Appendix \ref{app:obj_laplace},
    for Additive Independent Cauchy Noise (AICN) in Appendix \ref{app:obj_cauchy}  
    and give performance results for the same.
	\section{Results} \label{sec:results}
In this section, we describe and report results based on simulation studies to validate
the analysis and design of the proposed method. We compare the proposed method with
existing works in both traditional and deep learning based approaches.
For the purpose of evaluation, we consider
three cases:
\begin{enumerate}
    \item 2 bit block with one complex channel use ($M=4,m=2$). This scheme is
            similar to the QPSK scheme which uses one constellation point in 
            complex channel plane to represent 2 bits.
    \item 4 bit block with two complex channel uses $(M=16,m=4)$.
    \item 8 bit block with four complex channel uses $(M=256,m=8)$.
\end{enumerate}
All the schemes are evaluated in both AWGN and RBF channel models. We compare the performance 
of trained models with traditional methods of QAM and Agrell sphere packing \cite{agrell2014database}
and deep learning based method proposed in \cite{o2017introduction}. For deep learning
based methods, $100$ models are trained and the results are reported.

We use two metrics to compare the capabilities of the schemes.
\begin{enumerate}
    \item Block Error Rate (BLER): The block error rate performance over a wide
            range of SNR of the schemes will show the usefulness of the schemes
            in delivering the information over the channel.
    \item Packing Density: Another metric to compare the efficiency of multiple 
            signaling methods is to compare the packing density of the transmit 
            signals over the dimensions specified by the number of channel uses. 
            Normalized second moment ($E_n$) of the transmit symbols $\textbf{z}$ 
            is defined as \cite{agrell2014database}
            \begin{align}
                E_n = \frac{1}{M} \frac{1}{d_{min}^2} \sum \limits_{i=1}^{M} 
                        \textbf{z}_i^T\textbf{z}_i,   \label{eqn:packing_density}
            \end{align}
            where $d_{min}^2 = \underset{i \neq j}{\min}(\textbf{z}_i-\textbf{z}_j)^T(\textbf{z}_i-\textbf{z}_j)$
            is the square of minimum euclidean distance between transmit points.
            This metric is insensitive to scaling and hence useful to compare packing 
            densities. Smaller the value of $E_n$, better the packing density.
\end{enumerate}

Please refer Appendix \ref{app:sim_setup} for more details about the
simulation setup and the training procedure.

\begin{figure*}[!h]
    \centering
    \begin{subfigure}{.33\textwidth}
        \resizebox{\textwidth}{!}{
            \pgfplotstableread[col sep = comma]{./data/output_awgn_best_bler_02x01.csv}\datatable
            \pgfplotsset{xmin=0, xmax=10, xtick={0,2,...,10}, ymin=1e-4, ymax=1e-0}
            \begin{tikzpicture}[thick,scale=0.8]
    \begin{semilogyaxis}[
        width=12cm,
        height=9cm,
        grid=major,
        xlabel={SNR(dB)},
        ylabel={BLER},
        xlabel style={at={(0.50,0.00)}, font=\Large},
        ylabel style={at={(0.00,0.50)}, font=\Large},
        legend pos=south west,
        legend cell align={left},
        legend style={fill opacity=0.6, 
                        draw=none, row sep=2.0pt,
                        text opacity=1.0, font=\Large}
        ]
        
        \addplot[color_ae, solid, line width=2pt, mark=triangle, mark size=3.0] 
            table [y=y_ae, x=x_ae, col sep=comma]{\datatable};
        \addlegendentry{\upshape \cite{o2017introduction}};

        \addplot[color_vae_awgn, solid, line width=2pt, mark=square, mark size=3.0] 
            table [y=y_vae_awgn, x=x_vae_awgn, col sep=comma]{\datatable};
        \addlegendentry{\upshape Proposed: Trained with (\ref{exp:obj_awgn_1hot})};

        \addplot[color_vae_rbf, solid, line width=2pt, mark=o, mark size=3.0] 
            table [y=y_vae_rbf, x=x_vae_rbf, col sep=comma]{\datatable};
        \addlegendentry{\upshape Proposed: Trained with (\ref{exp:obj_rbf_1hot})};

        \addplot[color_qam, dash dot dot, line width=3pt] 
            table [y=y_QAM, x=x_QAM, col sep=comma]{\datatable};
        \addlegendentry{\upshape QAM};

        \addplot[color_agrell, dashdotted, line width=2pt] 
            table [y=y_Agrell, x=x_Agrell, col sep=comma]{\datatable};
        \addlegendentry{\upshape Agrell \cite{agrell2014database}};
    \end{semilogyaxis}
\end{tikzpicture}
        }
        \caption{BLER of best model}
        \label{fig:bler_best_awgn_02x01}
    \end{subfigure}%
    \begin{subfigure}{.33\textwidth}
        \resizebox{\textwidth}{!}{
            \pgfplotstableread[col sep = comma]{./data/output_awgn_en_02x01.csv}\datatable
            \pgfplotsset{xmin=0.49, xmax=0.55, xtick={0.49,0.50,...,0.56}}
\begin{tikzpicture}[tight background]
    \begin{axis}[
        width=12cm,
        height=9cm,
        ymin=-0.05,
        ymax=+1.05,
        grid=major,
        xlabel={$E_n$},
        ylabel={CDF},
        xlabel style={at={(0.50,-0.00)}},
        ylabel style={at={(0.00,0.50)}},
        ytick={0.0,0.2,...,1.0},
        label style={font=\Large},
        legend pos=north west,
        legend cell align={left},
        legend style={fill opacity=0.6, 
                    draw=none, row sep=2.0pt,
                    text opacity=1.0, font=\Large}
        ]
        
        \addplot[color_ae, solid, line width=2pt,
            mark=triangle, mark size={3.0}, mark repeat=4, mark phase=1] 
            table [y=y_ae, x=x_ae, col sep=comma]{\datatable};
        \addlegendentry{\upshape \cite{o2017introduction}};

        \addplot[color_vae_awgn, solid, line width=2pt,
            mark=square, mark size={3.0}, mark repeat=4, mark phase=1] 
            table [y=y_vae_awgn, x=x_vae_awgn, col sep=comma]{\datatable};
        \addlegendentry{\upshape Proposed: Trained with (\ref{exp:obj_awgn_1hot})};

        \addplot[color_vae_rbf, solid, line width=2pt,
            mark=o, mark size={3.0}, mark repeat=4, mark phase=1] 
            table [y=y_vae_rbf, x=x_vae_rbf, col sep=comma]{\datatable};
        \addlegendentry{\upshape Proposed: Trained with (\ref{exp:obj_rbf_1hot})};

        \addplot[color_qam, dash dot dot, line width=3pt] 
            table [y=y_QAM, x=x_QAM, col sep=comma]{\datatable};
        \addlegendentry{\upshape QAM};

        \addplot[color_agrell, dashdotted, line width=3pt] 
            table [y=y_Agrell, x=x_Agrell, col sep=comma]{\datatable};
        \addlegendentry{\upshape Agrell \cite{agrell2014database}};

    \end{axis}
\end{tikzpicture}
        }
        \caption{Distribution of $E_n$}
        \label{fig:en_awgn_02x01}
    \end{subfigure}%
    \begin{subfigure}{.33\textwidth}
        \resizebox{\textwidth}{!}{
            \pgfplotstableread[col sep = comma]{./data/output_awgn_bler06dB_evo_02x01.csv}\datatable
            \pgfplotsset{ymin=4.50e-2, ymax=4.75e-2, ytick={0.0450,0.0455,...,0.0480}}
            \begin{tikzpicture}[thick]
    \begin{semilogyaxis}[
        width=12cm,
        height=9cm,
        xmin=0,
        xmax=3000,
        grid=major,
        xlabel={Epochs},
        ylabel={BLER},
        xlabel style={at={(0.50,0.00)}, font=\Large},
        ylabel style={at={(0.00,0.50)}, font=\Large},
        log ticks with fixed point,
        legend pos=north east,
        legend cell align={left},
        legend style={fill opacity=0.6, 
                        draw=none, row sep=2.0pt,
                        text opacity=1.0, font=\Large}
        ]
        
        \addplot[color_ae, solid, line width=2pt,
            mark=triangle, mark size={3.0}, mark repeat=2, mark phase=1]
            table [y=y_ae, x=x_ae, col sep=comma]{\datatable};
        \addlegendentry{\upshape \cite{o2017introduction}};

        \addplot[color_vae_awgn, solid, line width=2pt,
            mark=square, mark size={3.0}, mark repeat=2, mark phase=1]
            table [y=y_vae_awgn, x=x_vae_awgn, col sep=comma]{\datatable};
        \addlegendentry{\upshape Proposed: Trained with (\ref{exp:obj_awgn_1hot})};

        \addplot[color_vae_rbf, solid, line width=2pt,
            mark=o, mark size={3.0}, mark repeat=2, mark phase=1] 
            table [y=y_vae_rbf, x=x_vae_rbf, col sep=comma]{\datatable};
        \addlegendentry{\upshape Proposed: Trained with (\ref{exp:obj_rbf_1hot})};

        \addplot[color_qam, dash dot dot, line width=3pt] 
            table [y=y_QAM, x=x_QAM, col sep=comma]{\datatable};
        \addlegendentry{\upshape QAM};

        \addplot[color_agrell, dashdotted, line width=2pt] 
            table [y=y_Agrell, x=x_Agrell, col sep=comma]{\datatable};
        \addlegendentry{\upshape Agrell \cite{agrell2014database}};




    \end{semilogyaxis}
\end{tikzpicture}
        }
        \caption{Evolution of BLER at $6dB$}
        \label{fig:bler06_evo_awgn_02x01}
    \end{subfigure}%
    \caption{Results for model with $M = 4, m = 2$ in AWGN channel}
    \label{fig:results_awgn_02x01}
\end{figure*}
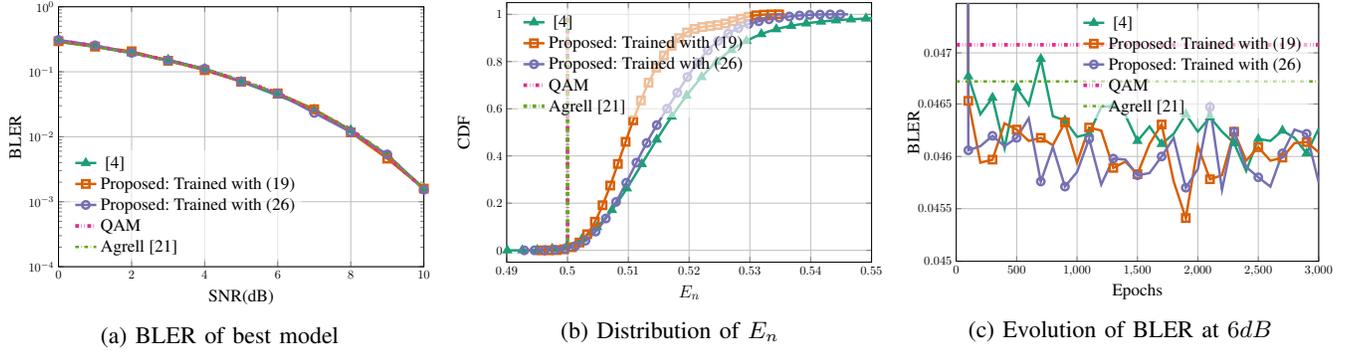

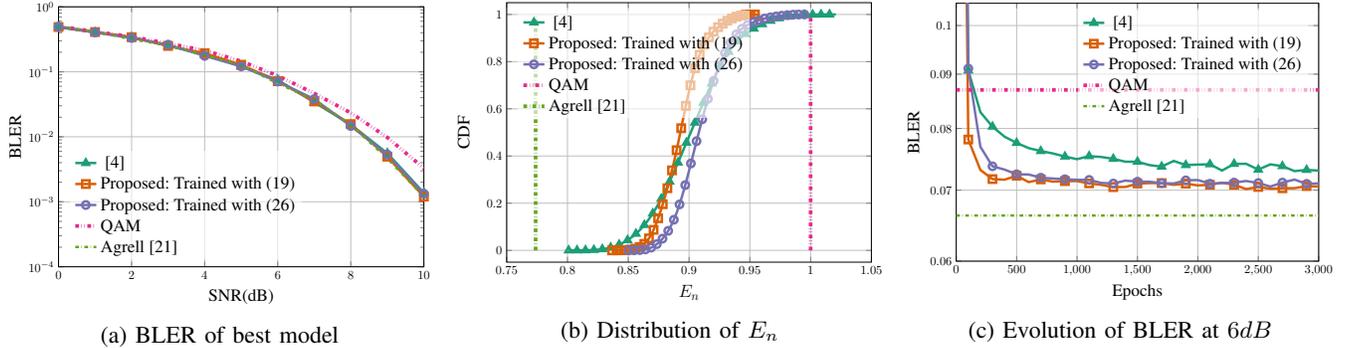
\begin{figure*}[!h]
    \centering
    \begin{subfigure}{.33\textwidth}
        \resizebox{\textwidth}{!}{
            \pgfplotstableread[col sep = comma]{./data/output_awgn_best_bler_04x02.csv}\datatable
            \pgfplotsset{xmin=0, xmax=10, xtick={0,2,...,10}, ymin=1e-4, ymax=1e-0}
            \begin{tikzpicture}[thick,scale=0.8]
    \begin{semilogyaxis}[
        width=12cm,
        height=9cm,
        grid=major,
        xlabel={SNR(dB)},
        ylabel={BLER},
        xlabel style={at={(0.50,0.00)}, font=\Large},
        ylabel style={at={(0.00,0.50)}, font=\Large},
        legend pos=south west,
        legend cell align={left},
        legend style={fill opacity=0.6, 
                        draw=none, row sep=2.0pt,
                        text opacity=1.0, font=\Large}
        ]
        
        \addplot[color_ae, solid, line width=2pt, mark=triangle, mark size=3.0] 
            table [y=y_ae, x=x_ae, col sep=comma]{\datatable};
        \addlegendentry{\upshape \cite{o2017introduction}};

        \addplot[color_vae_awgn, solid, line width=2pt, mark=square, mark size=3.0] 
            table [y=y_vae_awgn, x=x_vae_awgn, col sep=comma]{\datatable};
        \addlegendentry{\upshape Proposed: Trained with (\ref{exp:obj_awgn_1hot})};

        \addplot[color_vae_rbf, solid, line width=2pt, mark=o, mark size=3.0] 
            table [y=y_vae_rbf, x=x_vae_rbf, col sep=comma]{\datatable};
        \addlegendentry{\upshape Proposed: Trained with (\ref{exp:obj_rbf_1hot})};

        \addplot[color_qam, dash dot dot, line width=3pt] 
            table [y=y_QAM, x=x_QAM, col sep=comma]{\datatable};
        \addlegendentry{\upshape QAM};

        \addplot[color_agrell, dashdotted, line width=2pt] 
            table [y=y_Agrell, x=x_Agrell, col sep=comma]{\datatable};
        \addlegendentry{\upshape Agrell \cite{agrell2014database}};
    \end{semilogyaxis}
\end{tikzpicture}
        }
        \caption{BLER of best model}
        \label{fig:bler_best_awgn_04x02}
    \end{subfigure}%
    \begin{subfigure}{.33\textwidth}
        \resizebox{\textwidth}{!}{
            \pgfplotstableread[col sep = comma]{./data/output_awgn_en_04x02.csv}\datatable
            \pgfplotsset{xmin=0.75, xmax=1.05, xtick={0.75,0.80,...,1.10}}
\begin{tikzpicture}[tight background]
    \begin{axis}[
        width=12cm,
        height=9cm,
        ymin=-0.05,
        ymax=+1.05,
        grid=major,
        xlabel={$E_n$},
        ylabel={CDF},
        xlabel style={at={(0.50,-0.00)}},
        ylabel style={at={(0.00,0.50)}},
        ytick={0.0,0.2,...,1.0},
        label style={font=\Large},
        legend pos=north west,
        legend cell align={left},
        legend style={fill opacity=0.6, 
                    draw=none, row sep=2.0pt,
                    text opacity=1.0, font=\Large}
        ]
        
        \addplot[color_ae, solid, line width=2pt,
            mark=triangle, mark size={3.0}, mark repeat=4, mark phase=1] 
            table [y=y_ae, x=x_ae, col sep=comma]{\datatable};
        \addlegendentry{\upshape \cite{o2017introduction}};

        \addplot[color_vae_awgn, solid, line width=2pt,
            mark=square, mark size={3.0}, mark repeat=4, mark phase=1] 
            table [y=y_vae_awgn, x=x_vae_awgn, col sep=comma]{\datatable};
        \addlegendentry{\upshape Proposed: Trained with (\ref{exp:obj_awgn_1hot})};

        \addplot[color_vae_rbf, solid, line width=2pt,
            mark=o, mark size={3.0}, mark repeat=4, mark phase=1] 
            table [y=y_vae_rbf, x=x_vae_rbf, col sep=comma]{\datatable};
        \addlegendentry{\upshape Proposed: Trained with (\ref{exp:obj_rbf_1hot})};

        \addplot[color_qam, dash dot dot, line width=3pt] 
            table [y=y_QAM, x=x_QAM, col sep=comma]{\datatable};
        \addlegendentry{\upshape QAM};

        \addplot[color_agrell, dashdotted, line width=3pt] 
            table [y=y_Agrell, x=x_Agrell, col sep=comma]{\datatable};
        \addlegendentry{\upshape Agrell \cite{agrell2014database}};

    \end{axis}
\end{tikzpicture}
        }
        \caption{Distribution of $E_n$}
        \label{fig:en_awgn_04x02}
    \end{subfigure}%
    \begin{subfigure}{.33\textwidth}
        \resizebox{\textwidth}{!}{
            \pgfplotstableread[col sep = comma]{./data/output_awgn_bler06dB_evo_04x02.csv}\datatable
            \pgfplotsset{ymin=0.060, ymax=0.105, ytick={0.060,0.070,...,0.105}}
            \begin{tikzpicture}[thick]
    \begin{semilogyaxis}[
        width=12cm,
        height=9cm,
        xmin=0,
        xmax=3000,
        grid=major,
        xlabel={Epochs},
        ylabel={BLER},
        xlabel style={at={(0.50,0.00)}, font=\Large},
        ylabel style={at={(0.00,0.50)}, font=\Large},
        log ticks with fixed point,
        legend pos=north east,
        legend cell align={left},
        legend style={fill opacity=0.6, 
                        draw=none, row sep=2.0pt,
                        text opacity=1.0, font=\Large}
        ]
        
        \addplot[color_ae, solid, line width=2pt,
            mark=triangle, mark size={3.0}, mark repeat=2, mark phase=1]
            table [y=y_ae, x=x_ae, col sep=comma]{\datatable};
        \addlegendentry{\upshape \cite{o2017introduction}};

        \addplot[color_vae_awgn, solid, line width=2pt,
            mark=square, mark size={3.0}, mark repeat=2, mark phase=1]
            table [y=y_vae_awgn, x=x_vae_awgn, col sep=comma]{\datatable};
        \addlegendentry{\upshape Proposed: Trained with (\ref{exp:obj_awgn_1hot})};

        \addplot[color_vae_rbf, solid, line width=2pt,
            mark=o, mark size={3.0}, mark repeat=2, mark phase=1] 
            table [y=y_vae_rbf, x=x_vae_rbf, col sep=comma]{\datatable};
        \addlegendentry{\upshape Proposed: Trained with (\ref{exp:obj_rbf_1hot})};

        \addplot[color_qam, dash dot dot, line width=3pt] 
            table [y=y_QAM, x=x_QAM, col sep=comma]{\datatable};
        \addlegendentry{\upshape QAM};

        \addplot[color_agrell, dashdotted, line width=2pt] 
            table [y=y_Agrell, x=x_Agrell, col sep=comma]{\datatable};
        \addlegendentry{\upshape Agrell \cite{agrell2014database}};




    \end{semilogyaxis}
\end{tikzpicture}
        }
        \caption{Evolution of BLER at $6dB$}
        \label{fig:bler06_evo_awgn_04x02}
    \end{subfigure}%
    \caption{Results for model with $M = 16, m = 4$ in AWGN channel}
    \label{fig:results_awgn_04x02}
\end{figure*}

\subsection{DNN architecture}
    We consider a feedforward autoencoder architecture with three hidden dense layers 
    for encoder network and three hidden dense layers for decoder for all the experiments
    and both the DL methods under comparison for fairness. The network architecture 
    details are given in Table \ref{tab:network_arch}. 

    \begin{table}[!h]
        \centering
        \caption{Details of DNN architecture} \label{tab:network_arch}
        \begin{tabular}{clrl}
            \hline
            \multicolumn{1}{l}{}                                                             
                & \multicolumn{1}{c}{Layer Name}                      
                & \multicolumn{1}{c}{Size} 
                & \multicolumn{1}{c}{Activation Function}  
            \\ 
            \hline
            \hline
            \multirow{6}{*}{\begin{tabular}[c]{@{}c@{}}Transmitter\\ (Encoder)\end{tabular}} 
                & Input Layer                                         
                & $M$                      
                & -                                        \\
                
                & Hidden $E1$                                         
                & $64$                     
                & ReLU                                     \\
                
                & Hidden $E2$                                         
                & $32$                     
                & ReLU                                     \\
                
                & Hidden $E3$                                         
                & $16$                     
                & ReLU                                     \\
                
                & \multicolumn{1}{c}{\multirow{2}{*}{Transmit Layer}} 
                & \multirow{2}{*}{$m$}     
                    & Linear for Proposed           \\
                    & \multicolumn{1}{c}{}                                &                          
                    & Linear + BN for \cite{o2017introduction}        \\ 
            \hline
            
            \multirow{2}{*}{Channel}
                & \multirow{2}{*}{}
                & \multirow{2}{*}{}     
                    & (\ref{eqn:awgn_model}) for AWGN channel \\
                &&  & (\ref{eqn:rbf_model}) for RBF channel  \\ 
            \hline
            
            \multirow{4}{*}{\begin{tabular}[c]{@{}c@{}}Receiver \\ (Decoder)\end{tabular}}   
                & Hidden $D1$
                & $16$
                & ReLU\\

                & Hidden $D2$
                & $32$
                & ReLU\\

                & Hidden $D3$
                & $64$ & ReLU                                     \\
                                                                                            
                & Ouput Layer                                         
                & $M$                      
                & Softmax                                  \\
            \hline
        \end{tabular}
    \end{table}
    
    Selection of activation functions
    for the network layers impact both the quality of the solution as well as the
    convergence properties of the model. Traditional activation functions including
    sigmoid, tanh restrict the activations to be in the range of $[0,1]$ and $[-1,+1]$
    respectively with saturating effects near the boundaries. These saturation effects
    can hinder gradient propagation through the layers. Recent works applying
    deep learning for communication systems modeling advocate the use of advanced
    activation functions like Rectified Linear Units (ReLU) \cite{o2017introduction,dorner2018deep},
    Exponential Linear Units (ELU) \cite{aoudia2018model} etc. We use ReLU for activation
    at our hidden layers, linear activation at the output of the encoder network and
    a softmax layer for output of the decoder network.
    The works in \cite{o2017introduction,dorner2018deep,raj2018backpropagating,aoudia2018model} used a 
    Batch Normalization (BN) layer at the output of the transmitter to control the
    power of the transmitted constellation. If this layer is not included, the model
    will try to transmit at uncontrollably higher powers to minimize the cross-entropy
    loss. However, the objective functions presented in this work, (\ref{exp:obj_awgn_1hot})
    and (\ref{exp:obj_rbf_1hot}), includes an additional term to minimize the
    transmit power. Hence, the deep learning model is incentivized for doing power
    control at the learning phase and will control the constellation power 
    according to the noise it observed and reconstruction likelihood during training.
    We used $\sigma_0^2 = 1.0$ and $\sigma_n^2=0.1$ while training the proposed model.
    Adam optimizer \cite{kingma2014adam} with learning rate $0.01$, $\beta_1 = 0.99$
    and $\beta_2 = 0.999$ is used for training all models and each model is trained 
    for $3000$ epochs. The models using \cite{o2017introduction} are trained at 
    an SNR of $10dB$. 

    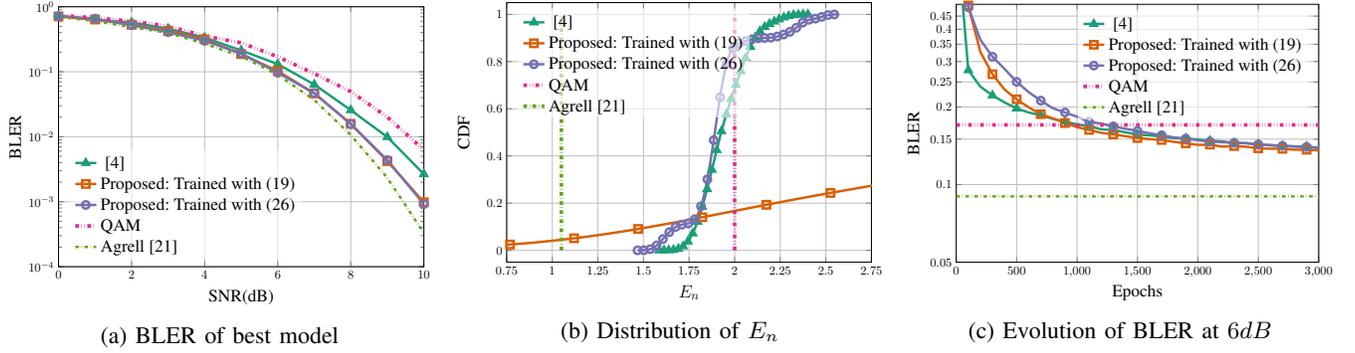
\begin{figure*}[!h]
        \centering
        \begin{subfigure}{.33\textwidth}
            \resizebox{\textwidth}{!}{
                \pgfplotstableread[col sep = comma]{./data/output_awgn_best_bler_08x04.csv}\datatable
                \pgfplotsset{xmin=0, xmax=10, xtick={0,2,...,10}, ymin=1e-4, ymax=1e-0}
                \begin{tikzpicture}[thick,scale=0.8]
    \begin{semilogyaxis}[
        width=12cm,
        height=9cm,
        grid=major,
        xlabel={SNR(dB)},
        ylabel={BLER},
        xlabel style={at={(0.50,0.00)}, font=\Large},
        ylabel style={at={(0.00,0.50)}, font=\Large},
        legend pos=south west,
        legend cell align={left},
        legend style={fill opacity=0.6, 
                        draw=none, row sep=2.0pt,
                        text opacity=1.0, font=\Large}
        ]
        
        \addplot[color_ae, solid, line width=2pt, mark=triangle, mark size=3.0] 
            table [y=y_ae, x=x_ae, col sep=comma]{\datatable};
        \addlegendentry{\upshape \cite{o2017introduction}};

        \addplot[color_vae_awgn, solid, line width=2pt, mark=square, mark size=3.0] 
            table [y=y_vae_awgn, x=x_vae_awgn, col sep=comma]{\datatable};
        \addlegendentry{\upshape Proposed: Trained with (\ref{exp:obj_awgn_1hot})};

        \addplot[color_vae_rbf, solid, line width=2pt, mark=o, mark size=3.0] 
            table [y=y_vae_rbf, x=x_vae_rbf, col sep=comma]{\datatable};
        \addlegendentry{\upshape Proposed: Trained with (\ref{exp:obj_rbf_1hot})};

        \addplot[color_qam, dash dot dot, line width=3pt] 
            table [y=y_QAM, x=x_QAM, col sep=comma]{\datatable};
        \addlegendentry{\upshape QAM};

        \addplot[color_agrell, dashdotted, line width=2pt] 
            table [y=y_Agrell, x=x_Agrell, col sep=comma]{\datatable};
        \addlegendentry{\upshape Agrell \cite{agrell2014database}};
    \end{semilogyaxis}
\end{tikzpicture}
            }
            \caption{BLER of best model}
            \label{fig:bler_best_awgn_08x04}
        \end{subfigure}%
        \begin{subfigure}{.33\textwidth}
            \resizebox{\textwidth}{!}{
                \pgfplotstableread[col sep = comma]{./data/output_awgn_en_08x04.csv}\datatable
                \pgfplotsset{xmin=0.75, xmax=2.75, xtick={0.75,1.00,...,2.75}}
\begin{tikzpicture}[tight background]
    \begin{axis}[
        width=12cm,
        height=9cm,
        ymin=-0.05,
        ymax=+1.05,
        grid=major,
        xlabel={$E_n$},
        ylabel={CDF},
        xlabel style={at={(0.50,-0.00)}},
        ylabel style={at={(0.00,0.50)}},
        ytick={0.0,0.2,...,1.0},
        label style={font=\Large},
        legend pos=north west,
        legend cell align={left},
        legend style={fill opacity=0.6, 
                    draw=none, row sep=2.0pt,
                    text opacity=1.0, font=\Large}
        ]
        
        \addplot[color_ae, solid, line width=2pt,
            mark=triangle, mark size={3.0}, mark repeat=4, mark phase=1] 
            table [y=y_ae, x=x_ae, col sep=comma]{\datatable};
        \addlegendentry{\upshape \cite{o2017introduction}};

        \addplot[color_vae_awgn, solid, line width=2pt,
            mark=square, mark size={3.0}, mark repeat=4, mark phase=1] 
            table [y=y_vae_awgn, x=x_vae_awgn, col sep=comma]{\datatable};
        \addlegendentry{\upshape Proposed: Trained with (\ref{exp:obj_awgn_1hot})};

        \addplot[color_vae_rbf, solid, line width=2pt,
            mark=o, mark size={3.0}, mark repeat=4, mark phase=1] 
            table [y=y_vae_rbf, x=x_vae_rbf, col sep=comma]{\datatable};
        \addlegendentry{\upshape Proposed: Trained with (\ref{exp:obj_rbf_1hot})};

        \addplot[color_qam, dash dot dot, line width=3pt] 
            table [y=y_QAM, x=x_QAM, col sep=comma]{\datatable};
        \addlegendentry{\upshape QAM};

        \addplot[color_agrell, dashdotted, line width=3pt] 
            table [y=y_Agrell, x=x_Agrell, col sep=comma]{\datatable};
        \addlegendentry{\upshape Agrell \cite{agrell2014database}};

    \end{axis}
\end{tikzpicture}
            }
            \caption{Distribution of $E_n$}
            \label{fig:en_awgn_08x04}
        \end{subfigure}%
        \begin{subfigure}{.33\textwidth}
            \resizebox{\textwidth}{!}{
                \pgfplotstableread[col sep = comma]{./data/output_awgn_bler06dB_evo_08x04.csv}\datatable
                \pgfplotsset{ymin=0.050, ymax=0.50, ytick={0.050,0.100,...,0.500}}
                \begin{tikzpicture}[thick]
    \begin{semilogyaxis}[
        width=12cm,
        height=9cm,
        xmin=0,
        xmax=3000,
        grid=major,
        xlabel={Epochs},
        ylabel={BLER},
        xlabel style={at={(0.50,0.00)}, font=\Large},
        ylabel style={at={(0.00,0.50)}, font=\Large},
        log ticks with fixed point,
        legend pos=north east,
        legend cell align={left},
        legend style={fill opacity=0.6, 
                        draw=none, row sep=2.0pt,
                        text opacity=1.0, font=\Large}
        ]
        
        \addplot[color_ae, solid, line width=2pt,
            mark=triangle, mark size={3.0}, mark repeat=2, mark phase=1]
            table [y=y_ae, x=x_ae, col sep=comma]{\datatable};
        \addlegendentry{\upshape \cite{o2017introduction}};

        \addplot[color_vae_awgn, solid, line width=2pt,
            mark=square, mark size={3.0}, mark repeat=2, mark phase=1]
            table [y=y_vae_awgn, x=x_vae_awgn, col sep=comma]{\datatable};
        \addlegendentry{\upshape Proposed: Trained with (\ref{exp:obj_awgn_1hot})};

        \addplot[color_vae_rbf, solid, line width=2pt,
            mark=o, mark size={3.0}, mark repeat=2, mark phase=1] 
            table [y=y_vae_rbf, x=x_vae_rbf, col sep=comma]{\datatable};
        \addlegendentry{\upshape Proposed: Trained with (\ref{exp:obj_rbf_1hot})};

        \addplot[color_qam, dash dot dot, line width=3pt] 
            table [y=y_QAM, x=x_QAM, col sep=comma]{\datatable};
        \addlegendentry{\upshape QAM};

        \addplot[color_agrell, dashdotted, line width=2pt] 
            table [y=y_Agrell, x=x_Agrell, col sep=comma]{\datatable};
        \addlegendentry{\upshape Agrell \cite{agrell2014database}};




    \end{semilogyaxis}
\end{tikzpicture}
            }
            \caption{Evolution of BLER at $6dB$}
            \label{fig:bler06_evo_awgn_08x04}
        \end{subfigure}%
        \caption{Results for model with $M = 256, m = 8$ in AWGN channel}
        \label{fig:results_awgn_08x04}
    \end{figure*}

    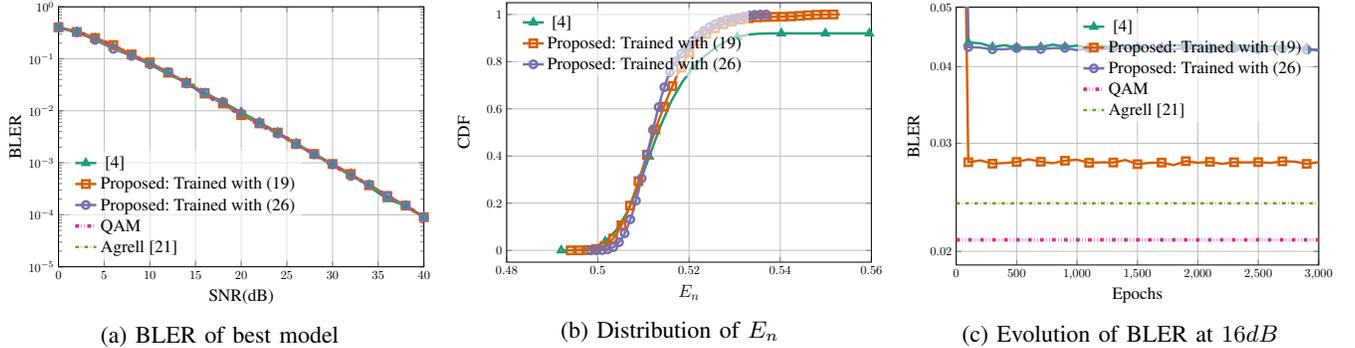
\begin{figure*}[!h]
        \centering
        \begin{subfigure}{.33\textwidth}
            \resizebox{\textwidth}{!}{
                \pgfplotstableread[col sep = comma]{./data/output_rbf_best_bler_02x01.csv}\datatable
                \pgfplotsset{xmin=0.0, xmax=40.0, xtick={0.0,5.0,...,40.0}, ymin=1e-5, ymax=1e-0}
                \begin{tikzpicture}[thick,scale=0.8]
    \begin{semilogyaxis}[
        width=12cm,
        height=9cm,
        grid=major,
        xlabel={SNR(dB)},
        ylabel={BLER},
        xlabel style={at={(0.50,0.00)}, font=\Large},
        ylabel style={at={(0.00,0.50)}, font=\Large},
        legend pos=south west,
        legend cell align={left},
        legend style={fill opacity=0.6, 
                        draw=none, row sep=2.0pt,
                        text opacity=1.0, font=\Large}
        ]
        
        \addplot[color_ae, solid, line width=2pt, mark=triangle, mark size=3.0] 
            table [y=y_ae, x=x_ae, col sep=comma]{\datatable};
        \addlegendentry{\upshape \cite{o2017introduction}};

        \addplot[color_vae_awgn, solid, line width=2pt, mark=square, mark size=3.0] 
            table [y=y_vae_awgn, x=x_vae_awgn, col sep=comma]{\datatable};
        \addlegendentry{\upshape Proposed: Trained with (\ref{exp:obj_awgn_1hot})};

        \addplot[color_vae_rbf, solid, line width=2pt, mark=o, mark size=3.0] 
            table [y=y_vae_rbf, x=x_vae_rbf, col sep=comma]{\datatable};
        \addlegendentry{\upshape Proposed: Trained with (\ref{exp:obj_rbf_1hot})};

        \addplot[color_qam, dash dot dot, line width=3pt] 
            table [y=y_QAM, x=x_QAM, col sep=comma]{\datatable};
        \addlegendentry{\upshape QAM};

        \addplot[color_agrell, dashdotted, line width=2pt] 
            table [y=y_Agrell, x=x_Agrell, col sep=comma]{\datatable};
        \addlegendentry{\upshape Agrell \cite{agrell2014database}};
    \end{semilogyaxis}
\end{tikzpicture}
            }
            \caption{BLER of best model}
            \label{fig:bler_best_rbf_02x01}
        \end{subfigure}%
        \begin{subfigure}{.33\textwidth}
            \resizebox{\textwidth}{!}{
                \pgfplotstableread[col sep = comma]{./data/output_rbf_en_02x01.csv}\datatable
                \pgfplotsset{xmin=0.48, xmax=0.56, xtick={0.48,0.50,...,0.57}}
    \begin{tikzpicture}[tight background]
        \begin{axis}[
            width=12cm,
            height=9cm,
            ymin=-0.05,
            ymax=+1.05,
            grid=major,
            xlabel={$E_n$},
            ylabel={CDF},
            xlabel style={at={(0.50,0.00)}, font=\Large},
            ylabel style={at={(0.00,0.50)}, font=\Large},
            ytick={0.0,0.2,...,1.0},
            label style={font=\Large},
            legend pos=north west,
            legend cell align={left},
            legend style={fill opacity=0.6, 
                        draw=none, row sep=2.0pt,
                        text opacity=1.0, font=\Large}
            ]
            
            \addplot[color_ae, solid, line width=2pt,
                mark=triangle, mark size={3.0}, mark repeat=4, mark phase=1]  
                table [y=y_ae, x=x_ae, col sep=comma]{\datatable};
            \addlegendentry{\upshape \cite{o2017introduction}};
    
            \addplot[color_vae_awgn, solid, line width=2pt,
                mark=square, mark size={3.0}, mark repeat=4, mark phase=1] 
                table [y=y_vae_awgn, x=x_vae_awgn, col sep=comma]{\datatable};
            \addlegendentry{\upshape Proposed: Trained with (\ref{exp:obj_awgn_1hot})};
    
            \addplot[color_vae_rbf, solid, line width=2pt,
                mark=o, mark size={3.0}, mark repeat=4, mark phase=1] 
                table [y=y_vae_rbf, x=x_vae_rbf, col sep=comma]{\datatable};
            \addlegendentry{\upshape Proposed: Trained with (\ref{exp:obj_rbf_1hot})};
    
    
    
        \end{axis}
    \end{tikzpicture}
            }
            \caption{Distribution of $E_n$}
            \label{fig:en_rbf_02x01}
        \end{subfigure}%
        \begin{subfigure}{.33\textwidth}
            \resizebox{\textwidth}{!}{
                \pgfplotstableread[col sep = comma]{./data/output_rbf_bler16dB_evo_02x01.csv}\datatable
                \pgfplotsset{ymin=1.9e-2, ymax=5.0e-2, ytick={0.02,0.03,...,0.05}}
                \begin{tikzpicture}[thick]
    \begin{semilogyaxis}[
        width=12cm,
        height=9cm,
        xmin=0,
        xmax=3000,
        grid=major,
        xlabel={Epochs},
        ylabel={BLER},
        xlabel style={at={(0.50,0.00)}, font=\Large},
        ylabel style={at={(0.00,0.50)}, font=\Large},
        log ticks with fixed point,
        legend pos=north east,
        legend cell align={left},
        legend style={fill opacity=0.6, 
                        draw=none, row sep=2.0pt,
                        text opacity=1.0, font=\Large}
        ]
        
        \addplot[color_ae, solid, line width=2pt,
            mark=triangle, mark size={3.0}, mark repeat=2, mark phase=1]
            table [y=y_ae, x=x_ae, col sep=comma]{\datatable};
        \addlegendentry{\upshape \cite{o2017introduction}};

        \addplot[color_vae_awgn, solid, line width=2pt,
            mark=square, mark size={3.0}, mark repeat=2, mark phase=1]
            table [y=y_vae_awgn, x=x_vae_awgn, col sep=comma]{\datatable};
        \addlegendentry{\upshape Proposed: Trained with (\ref{exp:obj_awgn_1hot})};

        \addplot[color_vae_rbf, solid, line width=2pt,
            mark=o, mark size={3.0}, mark repeat=2, mark phase=1] 
            table [y=y_vae_rbf, x=x_vae_rbf, col sep=comma]{\datatable};
        \addlegendentry{\upshape Proposed: Trained with (\ref{exp:obj_rbf_1hot})};

        \addplot[color_qam, dash dot dot, line width=3pt] 
            table [y=y_QAM, x=x_QAM, col sep=comma]{\datatable};
        \addlegendentry{\upshape QAM};

        \addplot[color_agrell, dashdotted, line width=2pt] 
            table [y=y_Agrell, x=x_Agrell, col sep=comma]{\datatable};
        \addlegendentry{\upshape Agrell \cite{agrell2014database}};




    \end{semilogyaxis}
\end{tikzpicture}
            }
            \caption{Evolution of BLER at $16dB$}
            \label{fig:bler16_evo_rbf_02x01}
        \end{subfigure}%
        \caption{Results for model with $M = 4, m = 2$ in RBF channel}
        \label{fig:results_rbf_02x01}
    \end{figure*}

\subsection{Evaluation in AWGN channel}
    The proposed method is evaluated in AWGN channel model given by (\ref{eqn:awgn_model}).
    In this case, the objective function to optimize is given in (\ref{exp:obj_awgn_1hot}).
    However, in a practical scenario, we would like to train the model without
    any assumptions on the channel model. To cover this case, we also provide results 
    using the objective function developed assuming RBF channel (\ref{exp:obj_rbf_1hot}).
    The results for different configurations under test are given in 
    Fig. \ref{fig:results_awgn_02x01} - \ref{fig:results_awgn_08x04}.

    The BLER vs SNR performance of the models are given in Fig. \ref{fig:bler_best_awgn_02x01},
    Fig. \ref{fig:bler_best_awgn_04x02} and Fig. \ref{fig:bler_best_awgn_08x04}.
    Agrell \cite{agrell2014database} being the optimized sphere packing scheme found
    using search is able to perform better in all cases. Note that, in the case
    of one complex channel use, both Agrell and QAM scheme are the same.
    As the number of channel uses increases, the dimension of the sphere packing
    problem also increases, and it can be seen that the QAM scheme does not
    perform as good as the other methods in comparison and the gap between the
    performance of Agrell scheme and QAM scheme widens with an increase in the number of
    channel uses.

    In all the cases, we can see that deep learning methods perform
    better than traditional QAM methods and are able to perform very close to the
    optimized Agrell schemes. Even with deep learning models, the performance compared
    to Agrell scheme widens as the dimension increases. 
    Interestingly, both (\ref{exp:obj_awgn_1hot})
    and (\ref{exp:obj_rbf_1hot}) provides equally good BLER performance in AWGN
    channel.

    The distribution of surrogate metric for packing density ,$E_n$ given by (\ref{eqn:packing_density}),
    for the trained models are given in Fig. \ref{fig:en_awgn_02x01}, Fig. \ref{fig:en_awgn_04x02}
    and Fig. \ref{fig:en_awgn_08x04}.
    We use kernel density
    estimation to smoothen the empirical histogram for packing density. 
    In the
    case of single channel use ($M=4,m=2$), traditional QAM and Agrell schemes are
    the optimal sphere packing schemes (with $E_n = 0.5$) and DL methods are able
    to reach close to this. 
    In the case of higher dimensions (Fig. \ref{fig:en_awgn_08x04}) , we can observe that
    the proposed objective function (\ref{exp:obj_rbf_1hot})
    is able to produce models with better $E_n$ than traditional QAM approximately 
    $90\%$ of the instances while the procedure in \cite{o2017introduction} managed
    to produce such models only $70\%$ of the time.
    From all the results, we can conclude
    that even though (\ref{exp:obj_rbf_1hot}) is developed for RBF channel, it can
    used in AWGN channel as well.
    
    Interesting observations emerge when we analyze the evolution of BLER of the
    models during the training phase (Fig. \ref{fig:bler06_evo_awgn_02x01}, 
    \ref{fig:bler06_evo_awgn_04x02} and \ref{fig:bler06_evo_awgn_08x04}).
    It can be observed that, in lower dimensions, the proposed loss functions
    are able to train the models faster than the method in \cite{o2017introduction}, 
    achieving lower BLERs early in the training.
    However, at higher dimensions (Fig. \ref{fig:bler06_evo_awgn_08x04}), the proposed
    loss functions slightly lags behind the method in \cite{o2017introduction} but 
    is eventually able to provide better BLER. These advantages during the training
    phase can be conclusively attributed to the new loss function developed
    in this work.
    
    At high dimensions, usage of
    (\ref{exp:obj_awgn_1hot}) results in high variability of packing density among 
    trained models,
    as seen from Fig. \ref{fig:en_awgn_08x04}. Even though this presents a difficulty
    in using these method at high dimensions, objective discussed in (\ref{exp:obj_rbf_1hot})
    is able to train better models consistently when compared with \cite{o2017introduction}.

\subsection{Evaluation in RBF channel}

    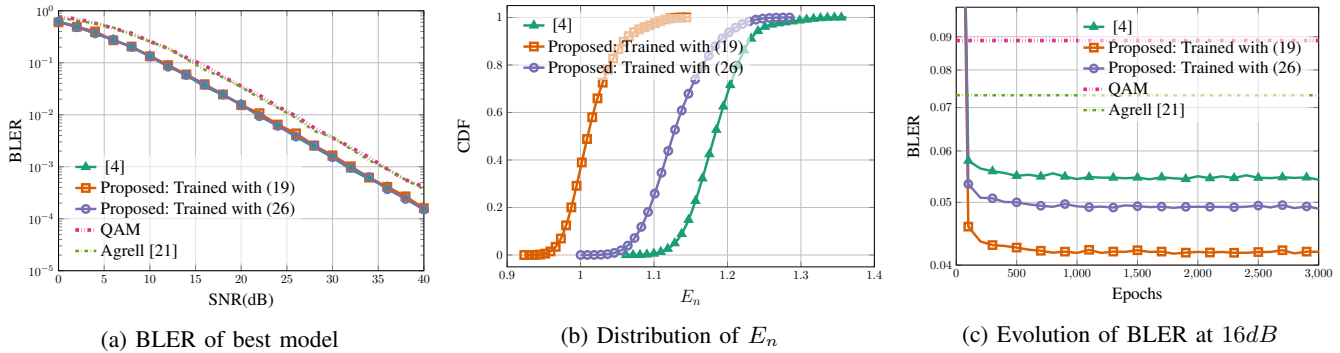
\begin{figure*}[!h]
        \centering
        \begin{subfigure}{.33\textwidth}
            \resizebox{\textwidth}{!}{
                \pgfplotstableread[col sep = comma]{./data/output_rbf_best_bler_04x02.csv}\datatable
                \pgfplotsset{xmin=0.0, xmax=40.0, xtick={0.0,5.0,...,40.0}, ymin=1e-5, ymax=1e-0}
                \begin{tikzpicture}[thick,scale=0.8]
    \begin{semilogyaxis}[
        width=12cm,
        height=9cm,
        grid=major,
        xlabel={SNR(dB)},
        ylabel={BLER},
        xlabel style={at={(0.50,0.00)}, font=\Large},
        ylabel style={at={(0.00,0.50)}, font=\Large},
        legend pos=south west,
        legend cell align={left},
        legend style={fill opacity=0.6, 
                        draw=none, row sep=2.0pt,
                        text opacity=1.0, font=\Large}
        ]
        
        \addplot[color_ae, solid, line width=2pt, mark=triangle, mark size=3.0] 
            table [y=y_ae, x=x_ae, col sep=comma]{\datatable};
        \addlegendentry{\upshape \cite{o2017introduction}};

        \addplot[color_vae_awgn, solid, line width=2pt, mark=square, mark size=3.0] 
            table [y=y_vae_awgn, x=x_vae_awgn, col sep=comma]{\datatable};
        \addlegendentry{\upshape Proposed: Trained with (\ref{exp:obj_awgn_1hot})};

        \addplot[color_vae_rbf, solid, line width=2pt, mark=o, mark size=3.0] 
            table [y=y_vae_rbf, x=x_vae_rbf, col sep=comma]{\datatable};
        \addlegendentry{\upshape Proposed: Trained with (\ref{exp:obj_rbf_1hot})};

        \addplot[color_qam, dash dot dot, line width=3pt] 
            table [y=y_QAM, x=x_QAM, col sep=comma]{\datatable};
        \addlegendentry{\upshape QAM};

        \addplot[color_agrell, dashdotted, line width=2pt] 
            table [y=y_Agrell, x=x_Agrell, col sep=comma]{\datatable};
        \addlegendentry{\upshape Agrell \cite{agrell2014database}};
    \end{semilogyaxis}
\end{tikzpicture}
            }
            \caption{BLER of best model}
            \label{fig:bler_best_rbf_04x02}
        \end{subfigure}%
        \begin{subfigure}{.33\textwidth}
            \resizebox{\textwidth}{!}{
                \pgfplotstableread[col sep = comma]{./data/output_rbf_en_04x02.csv}\datatable
                \pgfplotsset{xmin=0.90, xmax=1.40, xtick={0.90,1.00,...,1.45}}
    \begin{tikzpicture}[tight background]
        \begin{axis}[
            width=12cm,
            height=9cm,
            ymin=-0.05,
            ymax=+1.05,
            grid=major,
            xlabel={$E_n$},
            ylabel={CDF},
            xlabel style={at={(0.50,0.00)}, font=\Large},
            ylabel style={at={(0.00,0.50)}, font=\Large},
            ytick={0.0,0.2,...,1.0},
            label style={font=\Large},
            legend pos=north west,
            legend cell align={left},
            legend style={fill opacity=0.6, 
                        draw=none, row sep=2.0pt,
                        text opacity=1.0, font=\Large}
            ]
            
            \addplot[color_ae, solid, line width=2pt,
                mark=triangle, mark size={3.0}, mark repeat=4, mark phase=1]  
                table [y=y_ae, x=x_ae, col sep=comma]{\datatable};
            \addlegendentry{\upshape \cite{o2017introduction}};
    
            \addplot[color_vae_awgn, solid, line width=2pt,
                mark=square, mark size={3.0}, mark repeat=4, mark phase=1] 
                table [y=y_vae_awgn, x=x_vae_awgn, col sep=comma]{\datatable};
            \addlegendentry{\upshape Proposed: Trained with (\ref{exp:obj_awgn_1hot})};
    
            \addplot[color_vae_rbf, solid, line width=2pt,
                mark=o, mark size={3.0}, mark repeat=4, mark phase=1] 
                table [y=y_vae_rbf, x=x_vae_rbf, col sep=comma]{\datatable};
            \addlegendentry{\upshape Proposed: Trained with (\ref{exp:obj_rbf_1hot})};
    
    
    
        \end{axis}
    \end{tikzpicture}
            }
            \caption{Distribution of $E_n$}
            \label{fig:en_rbf_04x02}
        \end{subfigure}%
        \begin{subfigure}{.33\textwidth}
            \resizebox{\textwidth}{!}{
                \pgfplotstableread[col sep = comma]{./data/output_rbf_bler16dB_evo_04x02.csv}\datatable
                \pgfplotsset{ymin=4.0e-2, ymax=1.0e-1, ytick={0.04,0.05,...,0.10}}
                \begin{tikzpicture}[thick]
    \begin{semilogyaxis}[
        width=12cm,
        height=9cm,
        xmin=0,
        xmax=3000,
        grid=major,
        xlabel={Epochs},
        ylabel={BLER},
        xlabel style={at={(0.50,0.00)}, font=\Large},
        ylabel style={at={(0.00,0.50)}, font=\Large},
        log ticks with fixed point,
        legend pos=north east,
        legend cell align={left},
        legend style={fill opacity=0.6, 
                        draw=none, row sep=2.0pt,
                        text opacity=1.0, font=\Large}
        ]
        
        \addplot[color_ae, solid, line width=2pt,
            mark=triangle, mark size={3.0}, mark repeat=2, mark phase=1]
            table [y=y_ae, x=x_ae, col sep=comma]{\datatable};
        \addlegendentry{\upshape \cite{o2017introduction}};

        \addplot[color_vae_awgn, solid, line width=2pt,
            mark=square, mark size={3.0}, mark repeat=2, mark phase=1]
            table [y=y_vae_awgn, x=x_vae_awgn, col sep=comma]{\datatable};
        \addlegendentry{\upshape Proposed: Trained with (\ref{exp:obj_awgn_1hot})};

        \addplot[color_vae_rbf, solid, line width=2pt,
            mark=o, mark size={3.0}, mark repeat=2, mark phase=1] 
            table [y=y_vae_rbf, x=x_vae_rbf, col sep=comma]{\datatable};
        \addlegendentry{\upshape Proposed: Trained with (\ref{exp:obj_rbf_1hot})};

        \addplot[color_qam, dash dot dot, line width=3pt] 
            table [y=y_QAM, x=x_QAM, col sep=comma]{\datatable};
        \addlegendentry{\upshape QAM};

        \addplot[color_agrell, dashdotted, line width=2pt] 
            table [y=y_Agrell, x=x_Agrell, col sep=comma]{\datatable};
        \addlegendentry{\upshape Agrell \cite{agrell2014database}};




    \end{semilogyaxis}
\end{tikzpicture}
            }
            \caption{Evolution of BLER at $16dB$}
            \label{fig:bler16_evo_rbf_04x02}
        \end{subfigure}%
        \caption{Results for model with $M = 16, m = 4$ in RBF channel}
        \label{fig:results_rbf_04x02}
    \end{figure*}

    \begin{figure*}[!h]
        \centering
        \begin{subfigure}{.33\textwidth}
            \resizebox{\textwidth}{!}{
                \pgfplotstableread[col sep = comma]{./data/output_rbf_best_bler_08x04.csv}\datatable
                \pgfplotsset{xmin=0.0, xmax=40.0, xtick={0.0,5.0,...,40.0}, ymin=1e-5, ymax=1e-0}
                \begin{tikzpicture}[thick,scale=0.8]
    \begin{semilogyaxis}[
        width=12cm,
        height=9cm,
        grid=major,
        xlabel={SNR(dB)},
        ylabel={BLER},
        xlabel style={at={(0.50,0.00)}, font=\Large},
        ylabel style={at={(0.00,0.50)}, font=\Large},
        legend pos=south west,
        legend cell align={left},
        legend style={fill opacity=0.6, 
                        draw=none, row sep=2.0pt,
                        text opacity=1.0, font=\Large}
        ]
        
        \addplot[color_ae, solid, line width=2pt, mark=triangle, mark size=3.0] 
            table [y=y_ae, x=x_ae, col sep=comma]{\datatable};
        \addlegendentry{\upshape \cite{o2017introduction}};

        \addplot[color_vae_awgn, solid, line width=2pt, mark=square, mark size=3.0] 
            table [y=y_vae_awgn, x=x_vae_awgn, col sep=comma]{\datatable};
        \addlegendentry{\upshape Proposed: Trained with (\ref{exp:obj_awgn_1hot})};

        \addplot[color_vae_rbf, solid, line width=2pt, mark=o, mark size=3.0] 
            table [y=y_vae_rbf, x=x_vae_rbf, col sep=comma]{\datatable};
        \addlegendentry{\upshape Proposed: Trained with (\ref{exp:obj_rbf_1hot})};

        \addplot[color_qam, dash dot dot, line width=3pt] 
            table [y=y_QAM, x=x_QAM, col sep=comma]{\datatable};
        \addlegendentry{\upshape QAM};

        \addplot[color_agrell, dashdotted, line width=2pt] 
            table [y=y_Agrell, x=x_Agrell, col sep=comma]{\datatable};
        \addlegendentry{\upshape Agrell \cite{agrell2014database}};
    \end{semilogyaxis}
\end{tikzpicture}
            }
            \caption{BLER of best model}
            \label{fig:bler_best_rbf_08x04}
        \end{subfigure}%
        \begin{subfigure}{.33\textwidth}
            \resizebox{\textwidth}{!}{
                \pgfplotstableread[col sep = comma]{./data/output_rbf_en_08x04.csv}\datatable
                \pgfplotsset{xmin=1.25, xmax=3.00, xtick={1.00,1.25,...,3.50}}
    \begin{tikzpicture}[tight background]
        \begin{axis}[
            width=12cm,
            height=9cm,
            ymin=-0.05,
            ymax=+1.05,
            grid=major,
            xlabel={$E_n$},
            ylabel={CDF},
            xlabel style={at={(0.50,0.00)}, font=\Large},
            ylabel style={at={(0.00,0.50)}, font=\Large},
            ytick={0.0,0.2,...,1.0},
            label style={font=\Large},
            legend pos=north west,
            legend cell align={left},
            legend style={fill opacity=0.6, 
                        draw=none, row sep=2.0pt,
                        text opacity=1.0, font=\Large}
            ]
            
            \addplot[color_ae, solid, line width=2pt,
                mark=triangle, mark size={3.0}, mark repeat=4, mark phase=1]  
                table [y=y_ae, x=x_ae, col sep=comma]{\datatable};
            \addlegendentry{\upshape \cite{o2017introduction}};
    
            \addplot[color_vae_awgn, solid, line width=2pt,
                mark=square, mark size={3.0}, mark repeat=4, mark phase=1] 
                table [y=y_vae_awgn, x=x_vae_awgn, col sep=comma]{\datatable};
            \addlegendentry{\upshape Proposed: Trained with (\ref{exp:obj_awgn_1hot})};
    
            \addplot[color_vae_rbf, solid, line width=2pt,
                mark=o, mark size={3.0}, mark repeat=4, mark phase=1] 
                table [y=y_vae_rbf, x=x_vae_rbf, col sep=comma]{\datatable};
            \addlegendentry{\upshape Proposed: Trained with (\ref{exp:obj_rbf_1hot})};
    
    
    
        \end{axis}
    \end{tikzpicture}
            }
            \caption{Distribution of $E_n$}
            \label{fig:en_rbf_08x04}
        \end{subfigure}%
        \begin{subfigure}{.33\textwidth}
            \resizebox{\textwidth}{!}{
                \pgfplotstableread[col sep = comma]{./data/output_rbf_bler16dB_evo_08x04.csv}\datatable
                \pgfplotsset{ymin=4.0e-2, ymax=5.0e-1, ytick={0.04,0.10,0.20,0.30,0.40,0.50}}
                \begin{tikzpicture}[thick]
    \begin{semilogyaxis}[
        width=12cm,
        height=9cm,
        xmin=0,
        xmax=3000,
        grid=major,
        xlabel={Epochs},
        ylabel={BLER},
        xlabel style={at={(0.50,0.00)}, font=\Large},
        ylabel style={at={(0.00,0.50)}, font=\Large},
        log ticks with fixed point,
        legend pos=north east,
        legend cell align={left},
        legend style={fill opacity=0.6, 
                        draw=none, row sep=2.0pt,
                        text opacity=1.0, font=\Large}
        ]
        
        \addplot[color_ae, solid, line width=2pt,
            mark=triangle, mark size={3.0}, mark repeat=2, mark phase=1]
            table [y=y_ae, x=x_ae, col sep=comma]{\datatable};
        \addlegendentry{\upshape \cite{o2017introduction}};

        \addplot[color_vae_awgn, solid, line width=2pt,
            mark=square, mark size={3.0}, mark repeat=2, mark phase=1]
            table [y=y_vae_awgn, x=x_vae_awgn, col sep=comma]{\datatable};
        \addlegendentry{\upshape Proposed: Trained with (\ref{exp:obj_awgn_1hot})};

        \addplot[color_vae_rbf, solid, line width=2pt,
            mark=o, mark size={3.0}, mark repeat=2, mark phase=1] 
            table [y=y_vae_rbf, x=x_vae_rbf, col sep=comma]{\datatable};
        \addlegendentry{\upshape Proposed: Trained with (\ref{exp:obj_rbf_1hot})};

        \addplot[color_qam, dash dot dot, line width=3pt] 
            table [y=y_QAM, x=x_QAM, col sep=comma]{\datatable};
        \addlegendentry{\upshape QAM};

        \addplot[color_agrell, dashdotted, line width=2pt] 
            table [y=y_Agrell, x=x_Agrell, col sep=comma]{\datatable};
        \addlegendentry{\upshape Agrell \cite{agrell2014database}};




    \end{semilogyaxis}
\end{tikzpicture}
            }
            \caption{Evolution of BLER at $16dB$}
            \label{fig:bler16_evo_rbf_08x04}
        \end{subfigure}%
        \caption{Results for model with $M = 256, m = 8$ in RBF channel}
        \label{fig:results_rbf_08x04}
    \end{figure*}
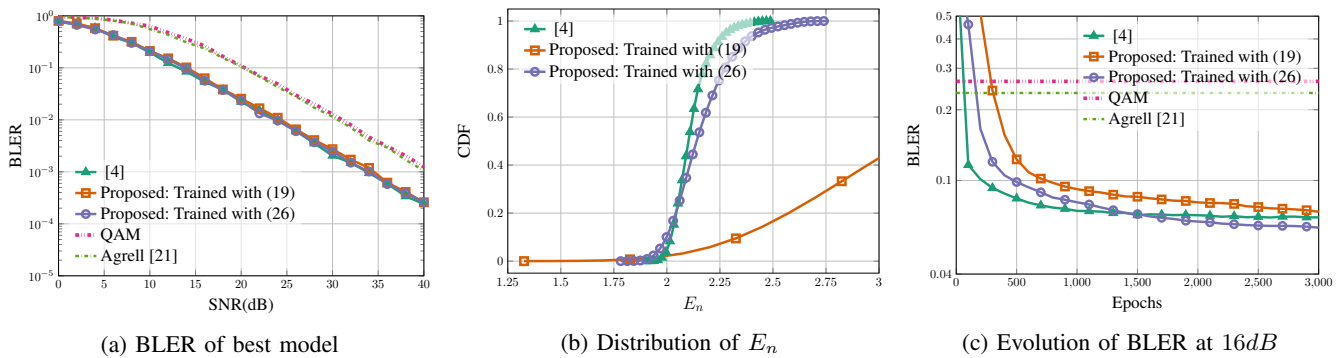

    For verifying the performance of the methods in the RBF channel model, the model
    described in (\ref{eqn:rbf_model}) is used. We provide the results of optimizing
    the DNN using both the objectives, (\ref{exp:obj_awgn_1hot}) and (\ref{exp:obj_rbf_1hot}),
    in Fig. \ref{fig:results_rbf_02x01} - \ref{fig:results_rbf_08x04}.

    We need to used pilot symbols to obtain an estimate of channel coefficient 
    $h$ and the equalization is done prior to decoding as done in \cite{ye2018channel,aoudia2018model}. 
    The estimate of $h$
    obtained from pilot symbols affects decoding performance through noisy
    equalization. We used the same power per component as the constellation points
    to transmit the pilot symbol such that both the pilot components and the symbol
    components in the block experience the same SNR during transmission. 

    Traditional QAM and Agrell schemes are not optimized for RBF channels. As
    the number of channel uses increases, we can see that the DL methods are able 
    to perform better than QAM and Agrell. The improvement in the case of DL methods
    can be attributed to the function approximation power of neural networks which
    learns to neutralize the effects of noisy channel equalization. Surprisingly,
    in RBF channel, the models trained with objective derived for AWGN model 
    (\ref{exp:obj_awgn_1hot}) is  able to give performance close to the models using
    (\ref{exp:obj_rbf_1hot}). 
    However, the difference when one uses (\ref{exp:obj_rbf_1hot})
    is visible in the packing density of the learned models. At higher dimension (Fig. \ref{fig:en_rbf_08x04}),
    (\ref{exp:obj_rbf_1hot}) is able to consistently produce better models when
    compared to (\ref{exp:obj_awgn_1hot}). Although the method in \cite{o2017introduction}
    is able to produce models with less variation at higher dimensions, in lower
    dimensions (Fig. \ref{fig:en_rbf_02x01} and Fig. \ref{fig:en_rbf_04x02}), it
    suffers with high variability. 
    The evolution of BLER of the trained models at $16dB$ is given in Fig. \ref{fig:bler16_evo_rbf_02x01},
    \ref{fig:bler16_evo_rbf_04x02} and \ref{fig:bler16_evo_rbf_08x04}. Interestingly,
    at low dimensions, the objective derived for AWGN channel model performs
    better that the objective for RBF model. Also, in all cases, the derived
    loss functions are able to provide a better BLER than the method in \cite{o2017introduction}.
    From all these results, we can conclude that
    using the objective (\ref{exp:obj_rbf_1hot}) derived for RBF channel model can 
    be expected to produce desired results consistently across different dimensions.

    Based on the above, it can be inferred that the proposed method for end to end 
    communication system design 
    \begin{enumerate}
        \item Provides a solution which accounts for noise corrupted latent codes
                with a theoretical backing.
        \item Consistently trains better models when compared to existing AE based
                methods.
    \end{enumerate}
    Now we investigate the impact of hyperparameter $\sigma_0^2$ and the input encoding
    for better insights into models and constellation labels.
    
\subsection{Effect of $\sigma_0^2$}
    We used a prior of $p(\hat{\textbf{z}}) = \mathcal{N}(\bm{0},\sigma_0^2 \bm{I})$
    with variance per component $\sigma_0^2$ during the derivation of objective
    functions (\ref{exp:obj_awgn_1hot}) and (\ref{exp:obj_rbf_1hot}). It can be
    easily seen from these objective functions that $\sigma_0^2$ affects the weight
    given to the transmit symbol power term $\sum \limits_{j=1}^{m} z_j^2$. When
    prior variance $\sigma_0^2$ is low, more weight is given to the transmit symbol
    power control term to reduce the transmit power and vice versa. 
    However, a very low value of $\sigma_0^2$
    will aggressively optimize the transmit power such that the constellations learned
    will have transmit power close to $0$. This affects the decoding process and increases
    the BLER. 
    
    Further, the numerical value of $\sigma_0^2$ is also related to the noise
    power $\sigma_n^2$. When noise power $\sigma_n^2$ is very high, the received symbol 
    $\hat{\textbf{z}}$ will be heavily distorted and hence a meaningful reconstruction 
    of the transmitted symbol is difficult. 
    This requires the models to transmit at higher
    power for learning to proceed which can be achieved by using higher numerical value
    for $\sigma_0^2$. 
    When noise power is low, the magnitude of $\sigma_0^2$ can be set to low value enabling one to use
    low power designs.
    Hence, one is required set the value of $\sigma_0^2$ proportional to $\sigma_n^2$
    with $\sigma_0^2 > \sigma_n^2$ to enable learning.

\subsection{Recovering Gray codes}

    \begin{figure}[!h]
        \centering
    \ifCLASSOPTIONonecolumn
        \begin{subfigure}{.30\columnwidth}
    \else
        \begin{subfigure}{.45\columnwidth}
    \fi
            \includegraphics[width=\linewidth]{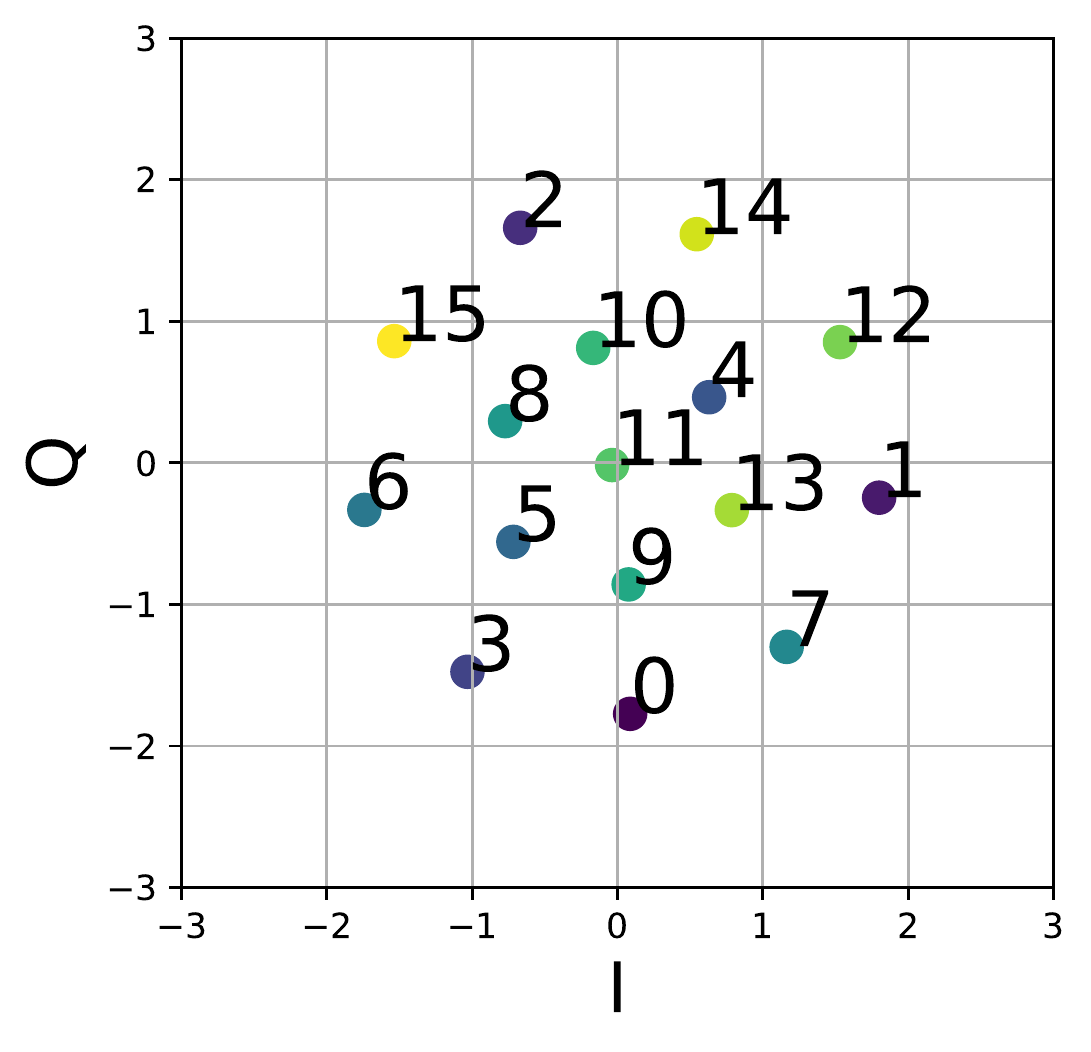}
            \caption{Sample 1}
            \label{fig:ae_const1}
        \end{subfigure}%
    \ifCLASSOPTIONonecolumn
        \begin{subfigure}{.30\columnwidth}
    \else
        \begin{subfigure}{.45\columnwidth}
    \fi
            \includegraphics[width=\linewidth]{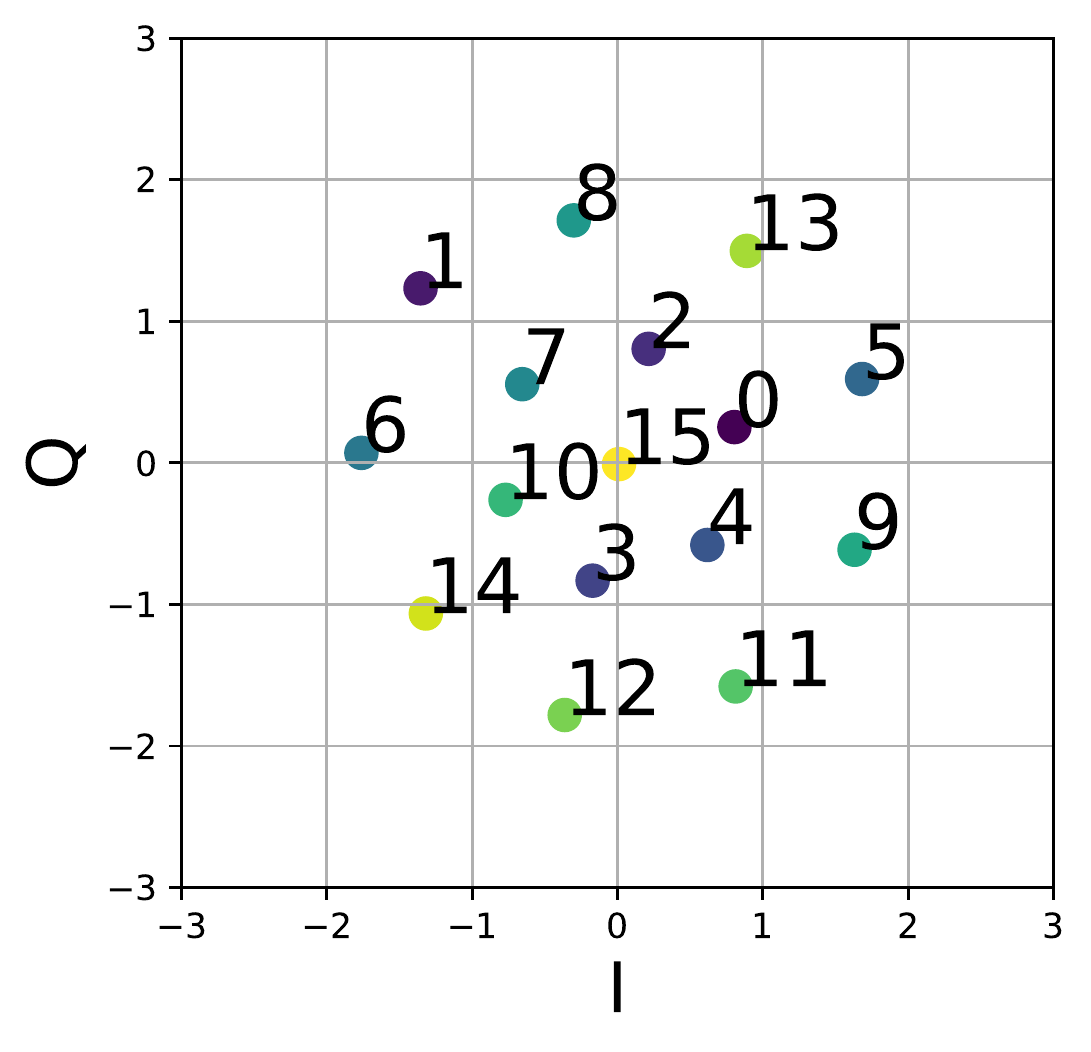}
            \caption{Sample 2}
            \label{fig:ae_const2}
        \end{subfigure}%
        \caption{Constellations trained for $M=16,m=2$ using \cite{o2017introduction}.}
        \label{fig:ae_code}
    \end{figure}

    In all the experiments discussed above, we used one-hot encoding to represent
    the symbols as done in previous works \cite{o2017introduction,dorner2018deep,
    raj2018backpropagating,aoudia2018model} for comparability. In order to study
    the structure of constellations, we trained models in AWGN channel for $M=16$ 
    and $m=2$ using the method in \cite{o2017introduction}\footnote{
    We chose $m=2$ for the simplicity of visualization as $m>2$ is difficult
    to visualize in 2D. Even though advanced methods like t-SNE can be used for high 
    dimensional visualization and analyzing clustering behavior as done in 
    \cite{o2017introduction,aoudia2018model},
    it is a projection to a 2D plane and may not efficiently covey the placements of
    points in high dimensional space which we are trying to analyze here.}.
    Two sample constellations learned by the method is given
    in Fig. \ref{fig:ae_code}. It can be observed that the symbols are well arranged in
    concentric circles maintaining sufficient distance between constellation points.
    This type of design is useful in optimizing BLER of the system. However, as 
    close-by symbols change by multiple bit positions, this may not be optimized
    way to design if the system requirement is to improve BER. This constellation
    characteristic is the effect of choosing one-hot encoding for representing symbols
    at both input and output
    as in one-hot encoding, there is no incentive for the model to place symbols
    with only one bit changes near to each other.

    However, by using the binary representation of the symbols and the reconstruction 
    likelihood introduced in (\ref{eqn:recon_bernoulli}), the models will be able 
    learn the concept of nearby symbols as the penalization forces all the bit positions
    to be correct. In this case, the objective function to train models in AWGN
    channel can be obtained by combining (\ref{eqn:recon_bernoulli}) and (\ref{eqn:klloss_awgn})
    and can be written as

    \begin{align}
        \max \left\{ 
            \sum \limits_{i = 1}^{d}
                \left( 
                    x_i \log \hat{x}_i +
                    (1-x_i) \log (1-\hat{x}_i)
                \right)
            - \frac{1}{2\sigma_0^2} \sum \limits_{j=1}^{m} z_j^2 \right\}.
            \label{exp:obj_awgn_binary}
    \end{align}

\begin{figure}[!h]
        \centering
\ifCLASSOPTIONonecolumn
            \begin{subfigure}{.30\columnwidth}
\else
            \begin{subfigure}{.45\columnwidth}
\fi
            \includegraphics[width=\linewidth]{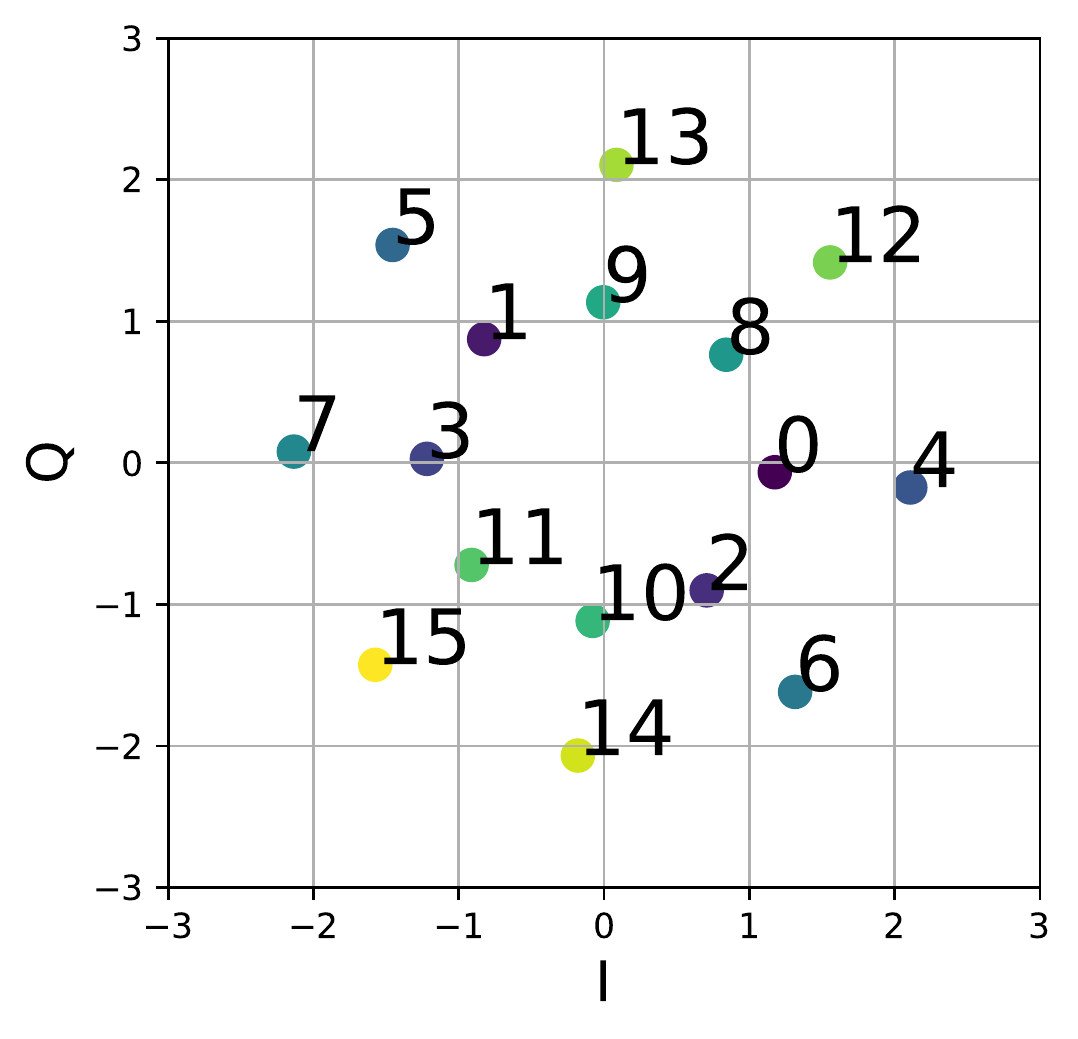}
            \caption{Sample 1}
            \label{fig:gray_const1}
        \end{subfigure}%
\ifCLASSOPTIONonecolumn
            \begin{subfigure}{.30\columnwidth}
\else
            \begin{subfigure}{.45\columnwidth}
\fi
            \includegraphics[width=\linewidth]{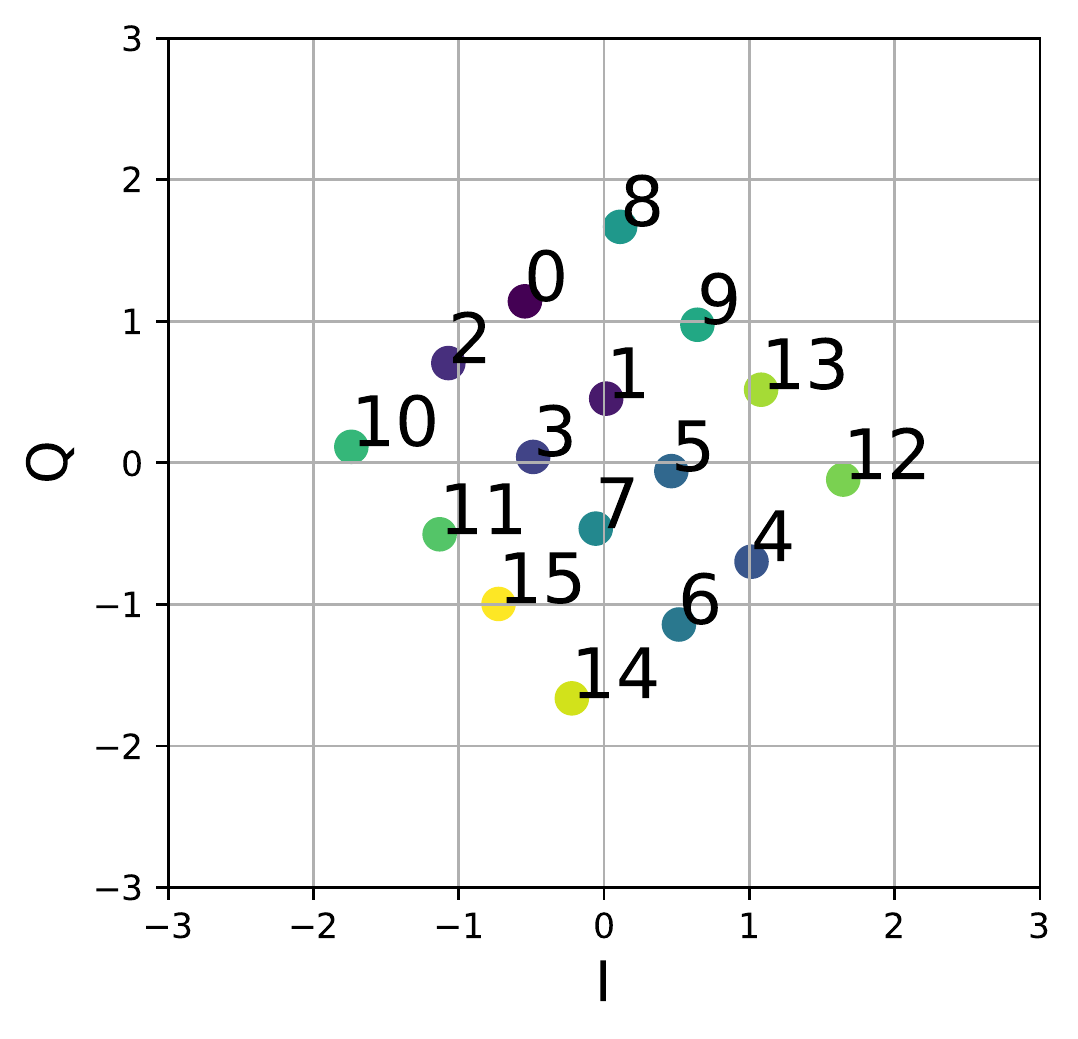}
            \caption{Sample 2}
            \label{fig:gray_const2}
        \end{subfigure}%
        \caption{Constellations trained for $M=16,m=2$ with (\ref{exp:obj_awgn_binary}).}
        \label{fig:gray_code}
    \end{figure}

    As the input layer dimension is now reduced from $16$ to $4$, we used a small network
    with hidden layers in encoder having $32$ and $16$ nodes, decoder having hidden
    layers with $16$ and $32$ nodes and finally an output layer of $4$ nodes with 
    sigmoid activation function. Training is done for $500$ epochs with other settings
    being similar to the one used in previous experiments. Sample constellations
    learned by this model is given in Fig. \ref{fig:gray_code}.

    From the constellations given in Fig \ref{fig:gray_code}, it can be easily observed
    that both the models \textit{learned} the concept of \textit{gray coding}.
    Symbols are placed in the constellation in such a way that near-by symbols vary
    by only one bit. After training multiple models, we observed that constellations
    with concentric circle structure as in Fig. \ref{fig:gray_const1} is the most
    commonly learned structure and the traditional grid-like structure as given in
    Fig. \ref{fig:gray_const2} occurs rarely. This shows that the loss function
    we use is having multiple local minima resulting in concentric structure and
    very few local minima resulting in a grid-like structure. 

    The use of explicit batchnormalization for constraining constellation energy
    in \cite{o2017introduction} results in one symbol being placed at point $(0,0)$
    as visible in Fig. \ref{fig:ae_code}. This may produce practical difficulties
    during transmission as a symbol close to $(0,0)$ is similar to no signal at 
    all. As the method proposed in this work includes constraining the constellation 
    energy into the objective function (\ref{exp:obj_awgn_binary}), this problem
    is not observed in the trained models (As seen in Fig. \ref{fig:gray_code}).
    The placement of a symbol at $(0,0)$ will result in constellation with
    center symbol differing in multiple bit positions from the symbols in first
    concentric circle and suffering a higher reconstruction likelihood with (\ref{eqn:recon_bernoulli}).
    Hence the models learn to avoid such a placement and instead places all symbols
    on concentric circles in the gray coding scheme.

\begin{figure}[!h]
        \centering
\ifCLASSOPTIONonecolumn
            \begin{subfigure}{.30\columnwidth}
\else
            \begin{subfigure}{.45\columnwidth}
\fi
            \includegraphics[width=\linewidth]{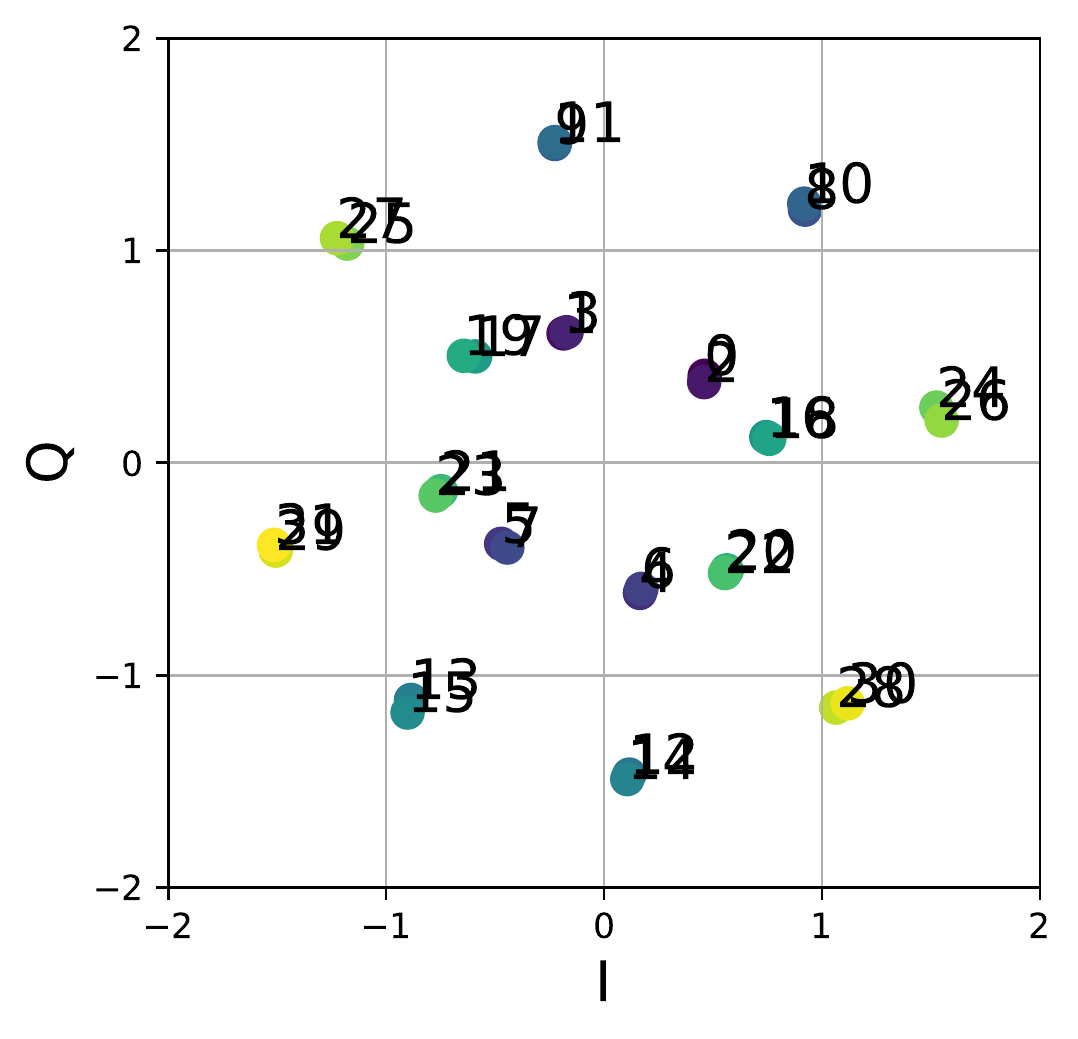}
            \caption{Learns to cheat ($\sigma_0^2 = 1.0$)}
            \label{fig:gray_const_32_cheat}
        \end{subfigure}
\ifCLASSOPTIONonecolumn
            \begin{subfigure}{.30\columnwidth}
\else
            \begin{subfigure}{.45\columnwidth}
\fi
            \includegraphics[width=\linewidth]{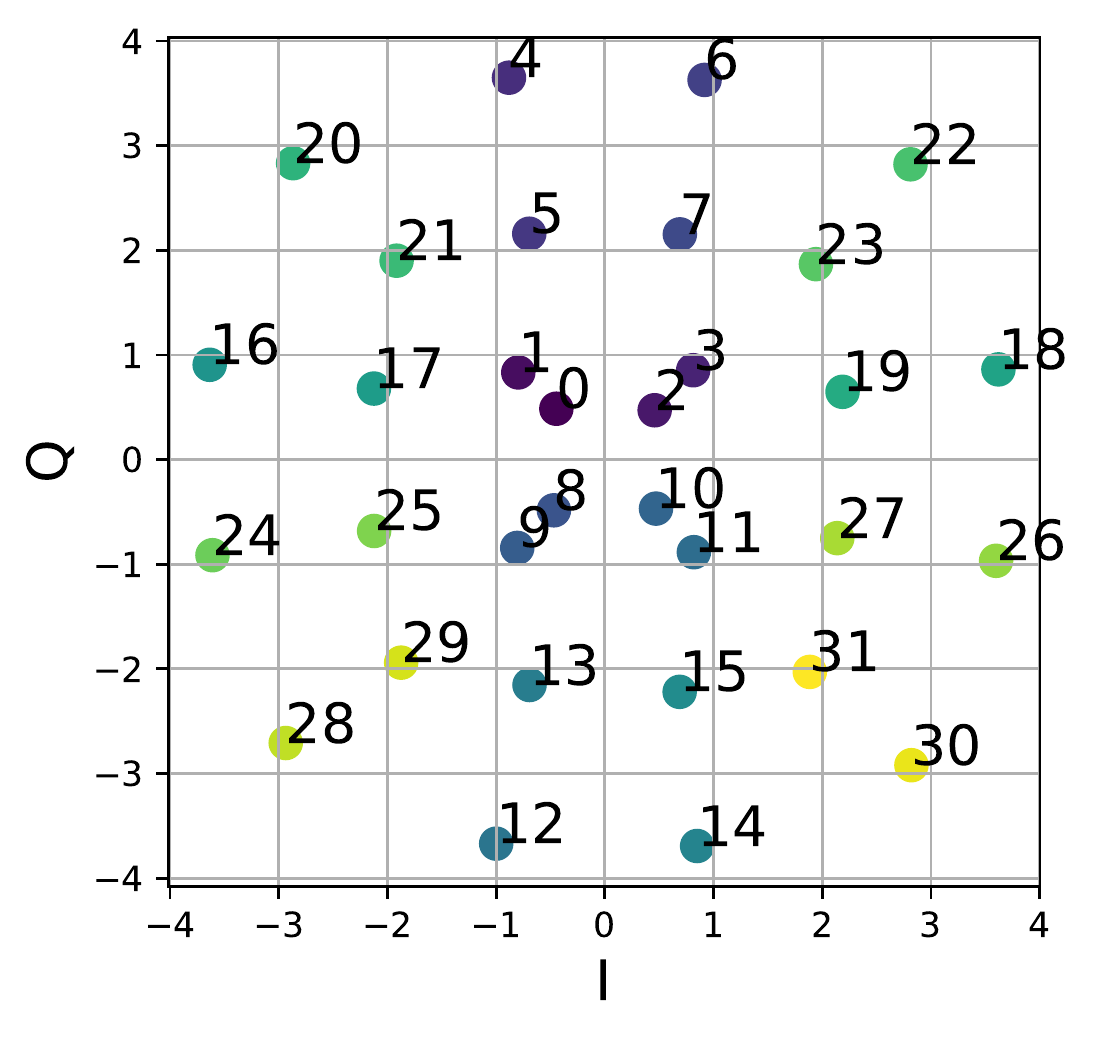}
            \caption{With $\sigma_0^2=5.0$}
            \label{fig:gray_const_32_good}
        \end{subfigure}
        \caption{Cheating behavior of model with variation in $\sigma_0^2$.}
        \label{fig:gray_code_32}
    \end{figure}

    Interestingly, when the number of symbols increased while keeping the $m=2$,
    the model learns to cheat the system by placing two symbols which 
    differ by only one bit top of each other and hence maintaining two concentric
    circles of the constellation but suffering a higher BLER. A sample constellation
    when the model is trained using $M=32$ is given in Fig. \ref{fig:gray_const_32_cheat}.

    In Fig. \ref{fig:gray_const_32_cheat}, the model learns to place symbols differing
    in second last bit (Eg: $(0,2), (29,31)$). We used a value of $\sigma_0^2 = 1.0$
    for this experiment. This cheating behavior can be attributed to the symbol
    energy control term in objective (\ref{exp:obj_awgn_binary}). 
    As discussed before, a low value of $\sigma_0^2$ will give more importance to 
    limiting the constellation transmit power and hence the model learns to place 
    symbols on top of each other while sacrificing reconstruction likelihood.
    During our experiments, we observed that increasing the value of $\sigma_0^2$
    also improved the BER performance (as expected, because of more spread out
    constellation). 
    By adjusting the value of $\sigma_0^2 = 5.0$, the model learns spread out the
    symbols while maintaining gray coding scheme as shown in Fig.\ref{fig:gray_const_32_good}.
    For this particular configuration under test, $M = 32, m =2$,
    we observed that increasing $\sigma_0^2$ beyond $5.0$ does not help in improving
    bit error rate.
    
    This behavior is observed when more bits are squeezed to transmit per
    channel use. It can be inferred that $\sigma_0^2$ acts as a \textit{honesty}
    parameter and when forcing the model to pack more bits per channel use,
    the model needs to have a high value for this parameter to avoid cheating
    behavior. As both $M$ and $\sigma_0^2$ are hyperparameters to be chosen during
    the system specification, this behavior can be easily handled by appropriately 
    setting the value of $\sigma_0^2$ at the design phase.

	\section{Concluding Remarks}
This work proposed a method to perform end to end modeling of communication systems
based on the principles of variational inference. Compared to the AE based systems
existing in the literature, the proposed method explicitly accounts for the noise 
corruption of latent codes (transmitted symbols). 
Further, unlike the AE based works which have a normalization layer leading
to hard constraints, we have adopted a soft constraint based approach.
The proposed
soft constraint approach enables models to explore better during the optimization process. 
Numerical simulation
results show that the proposed method is able to train faster, provides competitive BLER performance and
consistently better packing density compared to AE based designs. By modifying the loss
function, it is shown that the concepts of gray coding can be learned.
	
	\appendices

\section{Derivation of log-likelihood of data}   \label{app:loglikehood}
This derivation is based on \cite{kingma2013auto}. Noting that $p_\theta(\textbf{x})$
is constant with respect to $q_\phi(\cdot)$, we have
\begin{align*}
    \mathbb{E}_{p(\textbf{x})} \log p_\theta(\textbf{x}) 
        &= \mathbb{E}_{p(\textbf{x})} \mathbb{E}_{q_\phi(\hat{\textbf{z}}|\textbf{x})} \log p_\theta(\textbf{x}) \nonumber \\
        &= \mathbb{E}_{p(\textbf{x})} \mathbb{E}_{q_\phi(\hat{\textbf{z}}|\textbf{x})} \log\left( 
                \frac{p_\theta(\textbf{x},\hat{\textbf{z}})}
                     {p_\theta(\hat{\textbf{z}}|\textbf{x})}
            \right)  
            \nonumber \\
\ifCLASSOPTIONtwocolumn
        &= \mathbb{E}_{p(\textbf{x})} \mathbb{E}_{q_\phi(\hat{\textbf{z}}|\textbf{x})} \log\left( 
                \frac{p_\theta(\textbf{x},\hat{\textbf{z}})}
                     {q_\phi(\hat{\textbf{z}}|\textbf{x})}
            \right) + 
            \nonumber \\ & \qquad \qquad
            \mathbb{E}_{p(\textbf{x})} \mathbb{E}_{q_\phi(\hat{\textbf{z}}|\textbf{x})} \log\left( 
                \frac{q_\phi(\hat{\textbf{z}}|\textbf{x})}
                     {p_\theta(\hat{\textbf{z}}|\textbf{x})}
            \right) \nonumber \\
\else
        &= \mathbb{E}_{p(\textbf{x})} \mathbb{E}_{q_\phi(\hat{\textbf{z}}|\textbf{x})} \log\left( 
            \frac{p_\theta(\textbf{x},\hat{\textbf{z}})}
                {q_\phi(\hat{\textbf{z}}|\textbf{x})}
        \right) + 
        \mathbb{E}_{p(\textbf{x})} \mathbb{E}_{q_\phi(\hat{\textbf{z}}|\textbf{x})} \log\left( 
            \frac{q_\phi(\hat{\textbf{z}}|\textbf{x})}
                {p_\theta(\hat{\textbf{z}}|\textbf{x})}
        \right) \nonumber \\
\fi
        &= \mathcal{L}_{\theta,\phi}(\textbf{x}) + 
            \mathbb{E}_{p(\textbf{x})} \mathcal{D}_{KL}(q_\phi(\hat{\textbf{z}}|\textbf{x})||
                         p_\theta(\hat{\textbf{z}},\textbf{x})).
\end{align*}

\section{Derivation of objective function for AWGN channel} \label{app:awgn_obj}
    The KL-divergence between two normal distributions with $\bm{\mu} \in \mathbb{R}^m$ is given by
\ifCLASSOPTIONtwocolumn
    \begin{align}
        \mathcal{D}_{KL} &\left( \mathcal{N}(\bm{\mu}_1,\bm{\Sigma}_1) || 
                                \mathcal{N}(\bm{\mu}_2,\bm{\Sigma}_2) \right) \nonumber \\
            &= \frac{1}{2} \left[ 
                    tr(\bm{\Sigma}_2^{-1}\bm{\Sigma}_1) +
                    (\bm{\mu}_2-\bm{\mu}_1)^T \bm{\Sigma}_2^{-1}
                        (\bm{\mu}_2-\bm{\mu}_1) - \right. \nonumber \\
                    & \qquad \left. m + \log \frac{|\bm{\Sigma}_2|}{|\bm{\Sigma}_1|}
                \right].    \label{eqn:kldiv_mvnormal}
    \end{align}
\else
    \begin{align}
        \mathcal{D}_{KL} &\left( \mathcal{N}(\bm{\mu}_1,\bm{\Sigma}_1) || 
                                \mathcal{N}(\bm{\mu}_2,\bm{\Sigma}_2) \right)
            &= \frac{1}{2} \left[ 
                    tr(\bm{\Sigma}_2^{-1}\bm{\Sigma}_1) +
                    (\bm{\mu}_2-\bm{\mu}_1)^T \bm{\Sigma}_2^{-1}
                        (\bm{\mu}_2-\bm{\mu}_1) -  m 
                        + \log \frac{|\bm{\Sigma}_2|}{|\bm{\Sigma}_1|}
                \right].    \label{eqn:kldiv_mvnormal}
    \end{align}
\fi

    For AWGN model, we have $q_\phi(\hat{\textbf{z}|\textbf{x}}) = \mathcal{N}(\textbf{z},
    \sigma_n^2\bm{I}_m)$ and $p(\hat{\textbf{z}}) = \mathcal{N}(\bm{0}_m,\sigma_0^2\bm{I}_m)$. 
    Hence the KL Loss term in (\ref{eqn:elbo_split}) can be computed as
\ifCLASSOPTIONtwocolumn
    \begin{align}
        \mathcal{D}_{KL}&(q_\phi(\hat{\textbf{z}}|\textbf{x})||p(\hat{\textbf{z}})) 
            = \mathcal{D}_{KL} \left( \mathcal{N}(\textbf{z},\sigma_n^2\bm{I}_m) || 
                \mathcal{N}(\bm{0}_m,\sigma_0^2\bm{I}_m) \right) \nonumber \nonumber \\
            &= \frac{1}{2} \left[ Tr\left( (\sigma_0^2\bm{I}_m)^{-1} \sigma_n^2\bm{I}_m\right)
                + \textbf{z}^T (\sigma_0^2\bm{I}_m)^{-1} \textbf{z} \right. \nonumber \\
                    & \qquad \qquad \qquad \qquad \qquad \qquad - m \left.  
                    + \log \frac{|\sigma_0^2\bm{I}_m|}{|\sigma_n^2\bm{I}_m|} \right] \nonumber \\
            &= \frac{1}{2}\left[ m \frac{\sigma_n^2}{\sigma_0^2} 
                + \frac{1}{\sigma_0^2} \textbf{z}^T\textbf{z} - m
                + m \log \frac{\sigma_0^2}{\sigma_n^2}   \right] \nonumber \\
            &=  \frac{1}{2\sigma_0^2} \sum \limits_{j=1}^{m} z_j^2
                - \frac{m}{2} \left( 1 - \frac{\sigma_n^2}{\sigma_0^2} 
                    + \log \frac{\sigma_n^2}{\sigma_0^2} \right).
    \end{align}
\else
    \begin{align}
        \mathcal{D}_{KL}(q_\phi(\hat{\textbf{z}}|\textbf{x})||p(\hat{\textbf{z}})) 
            &= \mathcal{D}_{KL} \left( \mathcal{N}(\textbf{z},\sigma_n^2\bm{I}_m) || 
                \mathcal{N}(\bm{0}_m,\sigma_0^2\bm{I}_m) \right) \nonumber \nonumber \\
            &= \frac{1}{2} \left[ Tr\left( (\sigma_0^2\bm{I}_m)^{-1} \sigma_n^2\bm{I}_m\right)
                + \textbf{z}^T (\sigma_0^2\bm{I}_m)^{-1} \textbf{z} - m 
                    + \log \frac{|\sigma_0^2\bm{I}_m|}{|\sigma_n^2\bm{I}_m|} \right] \nonumber \\
            &= \frac{1}{2}\left[ m \frac{\sigma_n^2}{\sigma_0^2} 
                + \frac{1}{\sigma_0^2} \textbf{z}^T\textbf{z} - m
                + m \log \frac{\sigma_0^2}{\sigma_n^2}   \right] \nonumber \\
            &=  \frac{1}{2\sigma_0^2} \sum \limits_{j=1}^{m} z_j^2
                - \frac{m}{2} \left( 1 - \frac{\sigma_n^2}{\sigma_0^2} 
                    + \log \frac{\sigma_n^2}{\sigma_0^2} \right).
    \end{align}
\fi

\section{Derivation of objective function for RBF channel}   \label{app:rbf_obj}
    Noting that $\bm{\mu}_1 = \bm{\mu}_2 = \bm{0}_m$, $\bm{\Sigma}_1 = \frac{1}{2}\left( 
    \textbf{z}\textbf{z}^T - \textbf{Jz}\textbf{z}^T\textbf{J} \right) + \sigma_n^2 \bm{I}_m$
    and $\bm{\Sigma}_2 = \sigma_0^2 \bm{I}_m$, we have
\ifCLASSOPTIONtwocolumn
    \begin{align}
        \mathcal{D}_{KL}&(q_\phi(\hat{\textbf{z}}|\textbf{x})||p(\hat{\textbf{z}})) 
        \nonumber \\
            &= \mathcal{D}_{KL} \left( \mathcal{N}\left(\bm{0}_m,\frac{1}{2}\left( 
                \textbf{z}\textbf{z}^T - \textbf{Jz}\textbf{z}^T\textbf{J} \right) + \sigma_n^2\bm{I}_m\right) || 
                \right. \nonumber \\ & \left. \qquad \qquad \qquad \qquad \qquad \qquad
                \mathcal{N}(\bm{0}_m,\sigma_0^2\bm{I}_m) \right). \nonumber
    \end{align}
\else
    \begin{align}
        \mathcal{D}_{KL}&(q_\phi(\hat{\textbf{z}}|\textbf{x})||p(\hat{\textbf{z}})) 
            &= \mathcal{D}_{KL} \left( \mathcal{N}\left(\bm{0}_m,\frac{1}{2}\left( 
                \textbf{z}\textbf{z}^T - \textbf{Jz}\textbf{z}^T\textbf{J} \right) + \sigma_n^2\bm{I}_m\right) || 
                \mathcal{N}(\bm{0}_m,\sigma_0^2\bm{I}_m) \right). \nonumber
    \end{align}
\fi
    Simplifying,
    \begin{align}
        tr(\bm{\Sigma}_2^{-1}\bm{\Sigma}_1)
            &= tr\left( (\sigma_0^2 \bm{I}_m)^{-1} \left( \frac{1}{2}\left( 
                            \textbf{z}\textbf{z}^T - 
                                \textbf{Jz}\textbf{z}^T\textbf{J} \right) 
                        + \sigma_n^2\bm{I}_m \right) \right) \nonumber \\
            &= \frac{1}{2\sigma_0^2} tr \left( \textbf{z}\textbf{z}^T - 
                \textbf{Jz}\textbf{z}^T\textbf{J} \right) 
                    + \frac{\sigma_n^2}{\sigma_0^2} tr(\bm{I}_m) \nonumber \\
            &= \frac{1}{2\sigma_0^2} \left[ 
                    tr \left( \textbf{z}\textbf{z}^T \right) - 
                    tr \left( \textbf{Jz}\textbf{z}^T\textbf{J} \right)
                \right] + m \frac{\sigma_n^2}{\sigma_0^2} \nonumber \\
            &= \frac{1}{2\sigma_0^2} \left[  
                    tr \left( \textbf{z}\textbf{z}^T \right) - 
                    tr \left( \textbf{JJz}\textbf{z}^T \right)
                \right] + m \frac{\sigma_n^2}{\sigma_0^2} \nonumber \\
            &= \frac{1}{2\sigma_0^2} \left[  
                    tr \left( \textbf{z}\textbf{z}^T \right) +
                    tr \left( \textbf{z}\textbf{z}^T \right)
                \right] + m \frac{\sigma_n^2}{\sigma_0^2} \nonumber \\
            &= \frac{1}{\sigma_0^2} \sum\limits_{j=1}^{m} z_j^2 
                + m \frac{\sigma_n^2}{\sigma_0^2}.    \label{eqn:rbf_kl_t1}
    \end{align}
    Also, we have $\log |\bm{\Sigma}_2| = \log | \sigma_0^2 \bm{I}_m| 
    = m \log (\sigma_0^2)$ and 
    \begin{align}
        \log |\bm{\Sigma}_1|
            &= \log \left|\frac{1}{2}\left(\textbf{z}\textbf{z}^T - 
                        \textbf{Jz}\textbf{z}^T\textbf{J} \right) 
                            + \sigma_n^2\bm{I}_m\right| 
                            \nonumber \\
            &= \log \left| \frac{1}{2} \left(\textbf{z}\textbf{z}^T + 
                    \textbf{Jz}(\textbf{Jz})^T \right) 
                        + \sigma_n^2 \bm{I}_m \right| 
                            \quad(\because \textbf{J}^T = -\textbf{J}) \nonumber \\
            &= \log \left( \left(\sigma_n^2\right)^m \left( 1 + \frac{1}{\sigma_n^2} \textbf{z}\textbf{z}^T
                        + \frac{1}{4\sigma_n^4} \left( \textbf{z}\textbf{z}^T\right)^2
                    \right) \right) 
                    \nonumber \\
            &= \log \left( \left(\sigma_n^2\right)^m \left( 1 
                        + \frac{1}{2\sigma_n^2} \textbf{z}\textbf{z}^T \right)^2
                    \right) \nonumber \\
            &= m \log( \sigma_n^2 )  + 2 \log \left( 1 
                        + \frac{1}{2\sigma_n^2} \sum \limits_{j=1}^{m} z_j^2
                    \right).     \label{eqn:rbf_kl_t3}
    \end{align}
    Combining (\ref{eqn:kldiv_mvnormal}), (\ref{eqn:rbf_kl_t1}) and (\ref{eqn:rbf_kl_t3}),
    we get
\ifCLASSOPTIONtwocolumn
    \begin{align}
        \mathcal{D}_{KL} &\left( \mathcal{N}\left(\bm{0}_m,\frac{1}{2}\left( 
                \textbf{z}\textbf{z}^T - \textbf{Jz}\textbf{z}^T\textbf{J} \right) + \sigma_n^2\bm{I}_m\right) || 
                \mathcal{N}(\bm{0}_m,\sigma_0^2\bm{I}_m) \right) \nonumber \\
            &= \frac{1}{2} \left[ 
                \frac{1}{\sigma_0^2} \textbf{z}\textbf{z}^T 
                    + m \frac{\sigma_n^2}{\sigma_0^2}
                - m + m \log (\sigma_0^2) \right. \nonumber \\ 
                &\left. \qquad \qquad \qquad 
                - m \log( \sigma_n^2 )  
                    - 2 \log \left( 1 + 
                            \frac{1}{2\sigma_n^2} \textbf{z}\textbf{z}^T
                        \right)
                \right] \nonumber \\
            &= \frac{1}{2} \left[ \frac{1}{\sigma_0^2} \textbf{z}\textbf{z}^T
                    - m
                    + m \frac{\sigma_n^2}{\sigma_0^2}
                    - m \log \frac{\sigma_n^2}{\sigma_0^2} \right. \nonumber \\
                    & \left. \qquad \qquad \qquad \qquad \qquad
                    - 2 \log \left(  1 
                        + \frac{1}{2\sigma_n^2} \textbf{z}\textbf{z}^T
                    \right) \right]     \nonumber \\
            &= \frac{1}{2\sigma_0^2} \sum\limits_{j=1}^{m} z_j^2 
                    - \frac{m}{2} \left( 1 - \frac{\sigma_n^2}{\sigma_0^2} 
                        + \log \frac{\sigma_n^2}{\sigma_0^2}\right) \nonumber \\
                    & \qquad \qquad \qquad \qquad 
                    - \log \left( 1 + \frac{1}{2\sigma_n^2} \sum \limits_{j=1}^{m} z_j^2 \right).
    \end{align}
\else
    \begin{align}
        \mathcal{D}_{KL} &\left( \mathcal{N}\left(\bm{0}_m,\frac{1}{2}\left( 
                \textbf{z}\textbf{z}^T - \textbf{Jz}\textbf{z}^T\textbf{J} \right) + \sigma_n^2\bm{I}_m\right) || 
                \mathcal{N}(\bm{0}_m,\sigma_0^2\bm{I}_m) \right) \nonumber \\
            &= \frac{1}{2} \left[ 
                \frac{1}{\sigma_0^2} \textbf{z}\textbf{z}^T 
                    + m \frac{\sigma_n^2}{\sigma_0^2}
                - m + m \log (\sigma_0^2)  
                - m \log( \sigma_n^2 )  
                    - 2 \log \left( 1 + 
                            \frac{1}{2\sigma_n^2} \textbf{z}\textbf{z}^T
                        \right)
                \right] \nonumber \\
            &= \frac{1}{2} \left[ \frac{1}{\sigma_0^2} \textbf{z}\textbf{z}^T
                    - m
                    + m \frac{\sigma_n^2}{\sigma_0^2}
                    - m \log \frac{\sigma_n^2}{\sigma_0^2} 
                    - 2 \log \left(  1 
                        + \frac{1}{2\sigma_n^2} \textbf{z}\textbf{z}^T
                    \right) \right]     \nonumber \\
            &= \frac{1}{2\sigma_0^2} \sum\limits_{j=1}^{m} z_j^2 
                    - \frac{m}{2} \left( 1 - \frac{\sigma_n^2}{\sigma_0^2} 
                        + \log \frac{\sigma_n^2}{\sigma_0^2}\right) 
                    - \log \left( 1 + \frac{1}{2\sigma_n^2} \sum \limits_{j=1}^{m} z_j^2 \right).
    \end{align}
\fi

\section{Training models in Laplace noise environments} \label{app:obj_laplace}
    Laplace noise distribution is one of the popular noise models used in communication
    systems to capture the non-Gaussian impulsive behavior of signal corruption
    \cite{soury2014symbol,soury2015symbol,badarneh2015error}. Laplace noise model
    has been found useful modeling the signal corruption in cases of indoor and outdoor
    communications, ultra-wideband wireless systems, multi-user interference, etc (see
    \cite{soury2015symbol} and references therein). The probability density function
    of a Laplace random variable with mean $\mu$ and variance $2\sigma_n^2$ is defined as
    \cite{soury2015symbol}
    \begin{align}
        \mathcal{L}(x;\mu,\sigma_n) = \frac{1}{2\sigma_n} \exp \left(-\frac{|x-\mu|}{\sigma_n} \right).
    \end{align}
    The KL-divergence between two Laplace distributions can be derived as
\ifCLASSOPTIONtwocolumn
    \begin{align}
        \mathcal{D}_{KL}&(\mathcal{L}(x;\mu_1,\sigma_1)||\mathcal{L}(y;\mu_2,\sigma_2))
            = \frac{\sigma_1}{\sigma_2}
                \left(
                    \exp\left(- \frac{|\mu_1-\mu_2|}{\sigma_1}\right) 
                    \right. \nonumber \\ & \quad \left.
                    - \left( 1 - \frac{|\mu_1-\mu_2|}{\sigma_1}\right)
                    \right)
                + \frac{\sigma_1}{\sigma_2} - 1 
                - \log\left(\frac{\sigma_1}{\sigma_2}\right).
    \end{align}
\else
    \begin{align}
        \mathcal{D}_{KL}&(\mathcal{L}(x;\mu_1,\sigma_1)||\mathcal{L}(y;\mu_2,\sigma_2))
            = \frac{\sigma_1}{\sigma_2}
                \left(
                    \exp\left(- \frac{|\mu_1-\mu_2|}{\sigma_1}\right) 
                    - \left( 1 - \frac{|\mu_1-\mu_2|}{\sigma_1}\right)
                    \right)
                + \frac{\sigma_1}{\sigma_2} - 1 
                - \log\left(\frac{\sigma_1}{\sigma_2}\right).
    \end{align}
\fi

    Following the model in \cite{soury2015symbol}, the KL-loss term in (\ref{eqn:elbo_split})
    can be computed as
\ifCLASSOPTIONtwocolumn
    \begin{align}
        \mathcal{D}_{KL}&(q_\phi(\hat{\textbf{z}}|\textbf{x})||p(\hat{\textbf{z}})) 
            = \mathcal{D}_{KL}(\mathcal{L}(\hat{\textbf{z}};\textbf{z},\sigma_n^2)
                    ||\mathcal{L}(\hat{\textbf{z}};\bm{0},\sigma_n^2)) \nonumber \\
            &= \frac{\sigma_n}{\sigma_0} \sum \limits_{i=1}^{m} 
                            \exp \left( -\frac{|z_i|}{\sigma_n} \right)
                            - \frac{1}{\sigma_0} \sum \limits_{i=1}^{m} |z_i|
                            - m \log \frac{\sigma_n}{\sigma_0}
                            - m \nonumber \\
            &= \sum \limits_{i=1}^{m} \left( 
                    \frac{\sigma_n}{\sigma_0} \left(
                        \exp \left( - \frac{|z_i|}{\sigma_n} \right)
                            - \left( 1 - \frac{|z_i|}{m}\right)
                        \right) 
                        \right. \nonumber \\ &\left. \qquad \qquad \qquad \qquad \qquad
                        + \frac{\sigma_n}{\sigma_0}
                            - 1 - \log \frac{\sigma_n}{\sigma_0}
                    \right). \label{eqn:klloss_laplace_actual}
    \end{align}
\else
    \begin{align}
        \mathcal{D}_{KL}(q_\phi(\hat{\textbf{z}}|\textbf{x})||p(\hat{\textbf{z}})) 
            &= \mathcal{D}_{KL}(\mathcal{L}(\hat{\textbf{z}};\textbf{z},\sigma_n^2)
                        ||\mathcal{L}(\hat{\textbf{z}};\bm{0},\sigma_n^2)) \nonumber \\
            &= \frac{\sigma_n}{\sigma_0} \sum \limits_{i=1}^{m} 
                            \exp \left( -\frac{|z_i|}{\sigma_n} \right)
                            - \frac{1}{\sigma_0} \sum \limits_{i=1}^{m} |z_i|
                            - m \log \frac{\sigma_n}{\sigma_0}
                            - m \nonumber \\
                &= \sum \limits_{i=1}^{m} \left( 
                    \frac{\sigma_n}{\sigma_0} \left(
                        \exp \left( - \frac{|z_i|}{\sigma_n} \right)
                            - \left( 1 - \frac{|z_i|}{m}\right)
                        \right) 
                        + \frac{\sigma_n}{\sigma_0}
                            - 1 - \log \frac{\sigma_n}{\sigma_0}
                    \right). \label{eqn:klloss_laplace_actual}
    \end{align}
\fi
    This can be upper bounded using \cite[Lemma 2.5]{feldman2017calibrating},
    \begin{align}
        \mathcal{D}_{KL}(q_\phi(\hat{\textbf{z}}|\textbf{x})||p(\hat{\textbf{z}})) 
            &\leq \frac{\textbf{z}^T\textbf{z}}{2 \sigma_n \sigma_0}
                            + \frac{m}{7} \left(\frac{\sigma_0^2}{\sigma_n^2} - 1\right)^2
                                \frac{\sigma_n^2}{\sigma_0^2}. \label{eqn:klloss_laplace_ub}
    \end{align}
    In our experiments, we found that using (\ref{eqn:klloss_laplace_ub}) instead
    of (\ref{eqn:klloss_laplace_actual}) helps the model to learn fast. We suspect
    this is because of the $L_1$ term in (\ref{eqn:klloss_laplace_actual}) making
    the loss surface difficult to optimize over, while the upper bound in (\ref{eqn:klloss_laplace_ub})
    creates a smooth loss surface. Hence the objective function to train the models
    under Laplace noise can be written as,
    \begin{align}
        \underset{\bm{\theta}_T,\bm{\theta}_R}{\max} \left\{ 
            \sum \limits_{\textbf{x} \in \textbf{X}} \left(    
                \log(p_{\textbf{x}}) 
                - \frac{1}{2\sigma_n \sigma_0} \sum \limits_{j=1}^{m} z_j^2 
                \right)
            \right\}.
            \label{exp:obj_laplace_1hot}
    \end{align}

    \begin{figure}[!h]
        \centering
            \resizebox{0.75\linewidth}{!}{
                \pgfplotstableread[col sep = comma]{./data/output_lap_best_bler_04x02.csv}\datatable
                \pgfplotsset{xmin=0.0, xmax=10.0, xtick={0.0,2.0,...,10.0}, ymin=1e-3, ymax=1e-0}
                \begin{tikzpicture}[thick,scale=0.8]
    \begin{semilogyaxis}[
        width=12cm,
        height=9cm,
        grid=major,
        xlabel={SNR(dB)},
        ylabel={BLER},
        xlabel style={at={(0.50,0.00)}, font=\Large},
        ylabel style={at={(0.00,0.50)}, font=\Large},
        legend pos=south west,
        legend cell align={left},
        legend style={fill opacity=0.6, 
                        draw=none, row sep=2.0pt,
                        text opacity=1.0, font=\Large}
        ]
        



        \addplot[color_vae_lap, solid, line width=2pt, mark=o, mark size=3.0] 
            table [y=y_vae_lap, x=x_vae_lap, col sep=comma]{\datatable};
        \addlegendentry{\upshape Trained with (\ref{exp:obj_laplace_1hot})};

        \addplot[color_qam, dash dot dot, line width=3pt] 
            table [y=y_QAM, x=x_QAM, col sep=comma]{\datatable};
        \addlegendentry{\upshape QAM};

        \addplot[color_agrell, dashdotted, line width=2pt] 
            table [y=y_Agrell, x=x_Agrell, col sep=comma]{\datatable};
        \addlegendentry{\upshape Agrell \cite{agrell2014database}};
    \end{semilogyaxis}
\end{tikzpicture}
            }
            \caption{BLER Performance in Laplace noise for $M=16, m=4$.}
            \label{fig:output_lap_best_bler_04x02}
    \end{figure}
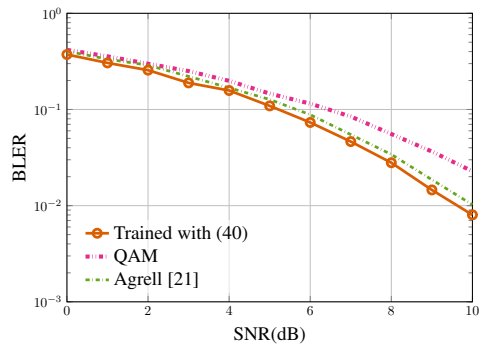

    The BLER performance of the proposed method with traditional methods for $M=16,m=4$ scheme
    is given in Fig. \ref{fig:output_lap_best_bler_04x02}. Agrell signalling scheme \cite{agrell2014database}
    is designed to be optimal for AWGN noise and in the case of additive Laplace noise,
    we can see that the proposed method is able to given better BLER performance.

\section{Training models in Cauchy noise environments.} \label{app:obj_cauchy}
    Cauchy noise distribution is a popular choice for modeling impulsive noise in 
    communication systems\cite{tsihrintzis1995incoherent,gulati2010statistics}. 
    However, the undefined nature of first and second moments of Additive Independent 
    Cauchy Noise (AICN) makes the analysis of such systems extremely difficult. The
    probability density function of AICN with location parameter $\delta$ and scale
    (dispersion) parameter $\gamma$ is given by \cite{fahs2014cauchy}:
    \begin{align}
        \mathcal{C}(x; \delta, \gamma)
            &= \frac{1}{\pi} \frac{\gamma}{\gamma^2 + (x-\delta)^2}.
    \end{align}
    The KL divergence between two Cauchy distributions can be shown as\cite{chyzak2019closed}:
    \begin{align}
        \mathcal{D}_{KL}(\mathcal{C}(x;\delta_1, \gamma_1)||\mathcal{C}(x;\delta_2,\gamma_2))
            &= \log \frac{(\gamma_1 + \gamma_2)^2 + (\delta_1 - \delta_2)^2}{4 \gamma_1 \gamma_2}.
    \end{align}

    In this experiment, we consider non-isometric Cauchy noise in $m$-dimensions
    and the KL-divergence between the received symbol (with dispersion $\gamma_n$
    per component) $q_{\phi}(\hat{\textbf{z}}; \textbf{z}, \gamma_n) = 
    \mathcal{C}(\hat{\textbf{z}}; \textbf{z}, \gamma_n)$ and $0$-location prior
    with $p(\hat{\textbf{z}}) = \mathcal{C}(\hat{\textbf{z}}; \textbf{0}, 
    \gamma_0)$ can be derived as:
    \begin{align}
        \mathcal{D}_{KL}(q_{\phi}(\hat{\textbf{z}}; \textbf{z}, \gamma_n) || 
        p(\hat{\textbf{z}})) 
            &= \sum \limits_{i = 1}^{m} 
                \log \frac{(\gamma_n + \gamma_0)^2 + z_i^2}{4 \gamma_n \gamma_0}.
            \label{eqn:kl_cauchy}
    \end{align}

    Combining this with the objective function derived in (\ref{eqn:elbo_split}) and using
    one-hot encoding, the objective for training models in Cauchy noise can be shown as
    \begin{align}
        \underset{\bm{\theta}_T,\bm{\theta}_R}{\max} \left\{ 
            \sum \limits_{\textbf{x} \in \textbf{X}} \left(    
                \log(p_{\textbf{x}}) 
                - \sum \limits_{i = 1}^{m} 
                    \log \frac{(\gamma_n + \gamma_0)^2 + z_i^2}{4 \gamma_n \gamma_0}
                \right)
            \right\}.
            \label{exp:obj_cauchy_1hot}
    \end{align}

    As the second moment of Cauchy distribution is undefined, the traditional definition
    of SNR is not applicable. Geometric SNR (G-SNR) is developed as an alternative to capture
    the noise strength and for a unit energy transmit symbol, G-SNR is defined as 
    \cite{gonzalez2006zero}
    \begin{align}
        G-SNR &= \frac{1}{2 C_g} \frac{1}{\gamma^2},
    \end{align}
    where $C_g \approx 1.78$.

    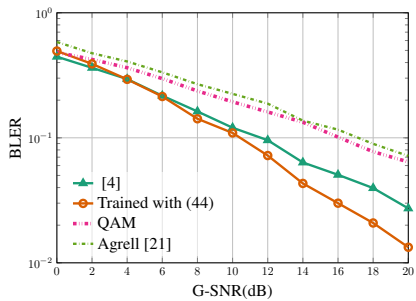
\begin{figure}[!h]
        \centering
            \resizebox{0.65\linewidth}{!}{
                \pgfplotstableread[col sep = comma]{./data/output_cau_best_bler_04x02.csv}\datatable
                \pgfplotsset{xmin=0.0, xmax=20.0, xtick={0.0,2.0,...,20.0}, ymin=1e-2, ymax=1e-0}
                \begin{tikzpicture}[thick,scale=0.8]
    \begin{semilogyaxis}[
        width=12cm,
        height=9cm,
        grid=major,
        xlabel={G-SNR(dB)},
        ylabel={BLER},
        xlabel style={at={(0.50,0.00)}, font=\Large},
        ylabel style={at={(0.00,0.50)}, font=\Large},
        legend pos=south west,
        legend cell align={left},
        legend style={fill opacity=0.6, 
                        draw=none, row sep=2.0pt,
                        text opacity=1.0, font=\Large}
        ]
        
        \addplot[color_ae, solid, line width=2pt, mark=triangle, mark size=3.0]
            table [y=y_ae, x=x_ae, col sep=comma]{\datatable};
        \addlegendentry{\upshape \cite{o2017introduction}};



        \addplot[color_vae_cau, solid, line width=2pt, mark=o, mark size=3.0] 
            table [y=y_vae_cau, x=x_vae_cau, col sep=comma]{\datatable};
        \addlegendentry{\upshape Trained with (\ref{exp:obj_cauchy_1hot})};

        \addplot[color_qam, dash dot dot, line width=3pt] 
            table [y=y_QAM, x=x_QAM, col sep=comma]{\datatable};
        \addlegendentry{\upshape QAM};

        \addplot[color_agrell, dashdotted, line width=2pt] 
            table [y=y_Agrell, x=x_Agrell, col sep=comma]{\datatable};
        \addlegendentry{\upshape Agrell \cite{agrell2014database}};
    \end{semilogyaxis}
\end{tikzpicture}
            }
            \caption{BLER Performance in Cauchy noise for $M=16, m=4$.}
            \label{fig:output_cau_best_bler_04x02}
    \end{figure}

    The BLER performance of proposed method when compared to traditional constellation
    designs in non-isometric channel is given in Fig. \ref{fig:output_cau_best_bler_04x02}
    for the case of $M = 16$ and $m = 4$. The models are trained with $\gamma_0 = 5.00$
    as we observed that lower value of prior dispersion adversely affect the learning 
    process.
    As Cauchy noise is very impulsive in nature, we
    can see that the BLER is also quite higher than that in other channels like AWGN, RBF
    etc for traditional constellations of QAM and Agrell. However, we can see that
    the proposed deep learning method is able to provide a huge margin in BLER over
    the traditional methods in Additive Independent Cauchy Noise channel.
    
\section{Details of Simulation Setup}   \label{app:sim_setup}
    We used Tensorflow-1.12 to implement deep learning models. All training 
    is done in a desktop-class computer with Intel Core $i7@2.4GHz$ CPU and $16GB$ RAM 
    and no GPU. For BLER results, the transmission of blocks are simulated until $500$
    block errors are observed, for both DL methods and traditional methods.

    While training in AWGN channel, even though the training set of symbols remained
    the same, we added different noise samples to each training point at each epoch.
    Similarly, for RBF channel, we used different values of channel coefficients $h$
    and noise samples at each epoch. This technique can reduce model overfitting
    as well as reduce the chances of getting stuck in saddle points.

    While training RBF models, we used equalization to condition the received symbol
    before feeding to the decoder network. We used a constant pilot symbol of $(1,1)$
    at transmission for equalization during the training phase. As the models trained 
    by the proposed methods do not guarantee constellation of specific energy,
    we need to appropriately modify the pilot energy during the testing phase. During 
    the testing phase, we maintained the per-component power of pilot to be equal to
    the average per component power of transmit symbols. This way, we can ensure
    that both pilot and data symbols experience the same SNR during testing. Pilot boosting
    can be used to improve that estimation of channel coefficients and hence BLER but 
    is out of the scope of this work.

\section{Training models in real channels}
    Since we assumed the knowledge of the channel and used a model-based simulation
    system, we were able to train the system with actual gradients. However, in 
    a real system, the channel impairments will be an unknown layer on the network
    and hence backpropagation of gradients from receiver to transmitter is not
    possible using traditional optimization techniques used by the deep learning community.
    Specific to the wireless communication domain, a few practical techniques are
    developed by the community to mitigate this problem and few of them are discussed
    below. This includes fine-tuning the receiver decoder with real channel \cite{dorner2018deep},
    using GANs\cite{goodfellow2014generative} to approximate the channel behavior
    \cite{ye2018channel,o2018physical}, approximating the channel gradients 
    \cite{raj2018backpropagating} by perturbation, perturbing the transmitter
    outputs \cite{aoudia2018model} etc.   
    We can replace the optimization objectives in these works
    with the objective function given in (\ref{exp:obj_general}) 
    and any of the following techniques can be used for model-free training with
    no further changes.
    
    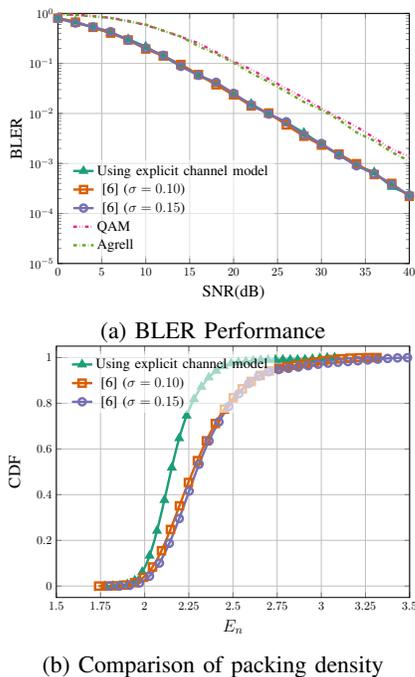
\begin{figure}[!h]
        \centering
\ifCLASSOPTIONonecolumn
        \begin{subfigure}{.40\columnwidth}
\else
        \begin{subfigure}{.65\columnwidth}
\fi
            \resizebox{\linewidth}{!}{
                \pgfplotstableread[col sep = comma]{./data/aoudia_output_rbf_best_bler_08x04.csv}\datatable
                \pgfplotsset{xmin=0.0, xmax=40.0, xtick={0.0,5.0,...,40.0}, ymin=1e-5, ymax=1e-0}
                \begin{tikzpicture}[thick,scale=0.8]
    \begin{semilogyaxis}[
        width=12cm,
        height=9cm,
        grid=major,
        xlabel={SNR(dB)},
        ylabel={BLER},
        xlabel style={at={(0.50,0.00)}, font=\Large},
        ylabel style={at={(0.00,0.50)}, font=\Large},
        legend pos=south west,
        legend cell align={left},
        legend style={fill opacity=0.6, 
                        draw=none, row sep=2.0pt,
                        text opacity=1.0, font=\large}
        ]
        
        \addplot[dark2_0, solid, line width=2pt, mark=triangle, mark size={3.0}] 
            table [y=y_adam, x=x_adam, col sep=comma]{\datatable};
        \addlegendentry{\upshape Using explicit channel model};

        \addplot[dark2_1, solid, line width=2pt, mark=square, mark size={3.0}]
            table [y=y_aoudia_s0010, x=x_aoudia_s0010, col sep=comma]{\datatable};
        \addlegendentry{\upshape \cite{aoudia2018model} ($\sigma = 0.10$)};

        \addplot[dark2_2, solid, line width=2pt, mark=o, mark size={3.0}]
            table [y=y_aoudia_s0015, x=x_aoudia_s0015, col sep=comma]{\datatable};
        \addlegendentry{\upshape \cite{aoudia2018model} ($\sigma = 0.15$)};

        \addplot[dark2_3, dash dot dot, line width=2pt]
            table [y=y_QAM, x=x_QAM, col sep=comma]{\datatable};
        \addlegendentry{\upshape QAM};

        \addplot[dark2_4, dashdotted, line width=2pt]
            table [y=y_Agrell, x=x_Agrell, col sep=comma]{\datatable};
        \addlegendentry{\upshape Agrell};
    \end{semilogyaxis}
\end{tikzpicture}
            }
            \caption{BLER Performance}
            \label{fig:aoudia_output_rbf_best_bler_08x04}
        \end{subfigure}
\ifCLASSOPTIONonecolumn
        \begin{subfigure}{.40\columnwidth}
\else
        \begin{subfigure}{.65\columnwidth}
\fi
            \resizebox{\linewidth}{!}{
                \pgfplotstableread[col sep = comma]{./data/aoudia_output_rbf_en_08x04.csv}\datatable
                \pgfplotsset{xmin=1.50, xmax=3.50, xtick={1.00,1.25,...,3.75}}
    \begin{tikzpicture}[tight background]
        \begin{axis}[
            width=12cm,
            height=9cm,
            ymin=-0.05,
            ymax=+1.05,
            grid=major,
            xlabel={$E_n$},
            ylabel={CDF},
            xlabel style={at={(0.50,0.00)}, font=\Large},
            ylabel style={at={(0.00,0.50)}, font=\Large},
            ytick={0.0,0.2,...,1.0},
            label style={font=\Large},
            legend pos=north west,
            legend cell align={left},
            legend style={fill opacity=0.6, 
                        draw=none, row sep=2.0pt,
                        text opacity=1.0, font=\large}
            ]
            
            \addplot[dark2_0, solid, line width=2pt,
                mark=triangle, mark size={3.0}, mark repeat=4, mark phase=1] 
                table [y=y_adam, x=x_adam, col sep=comma]{\datatable};
            \addlegendentry{\upshape Using explicit channel model};

            \addplot[dark2_1, solid, line width=2pt,
                mark=square, mark size={3.0}, mark repeat=4, mark phase=1] 
                table [y=y_aoudia_s0010, x=x_aoudia_s0010, col sep=comma]{\datatable};
            \addlegendentry{\upshape \cite{aoudia2018model} ($\sigma = 0.10$)};

            \addplot[dark2_2, solid, line width=2pt,
                mark=o, mark size={3.0}, mark repeat=4, mark phase=1]  
                table [y=y_aoudia_s0015, x=x_aoudia_s0015, col sep=comma]{\datatable};
            \addlegendentry{\upshape \cite{aoudia2018model} ($\sigma = 0.15$)};
    
        \end{axis}
    \end{tikzpicture}
            }
            \caption{Comparison of packing density}
            \label{fig:aoudia_output_rbf_en_08x04}
        \end{subfigure}%
        \caption{Comparison of models trained with RBF objective function
            and different training methods for $M = 256$ and $m = 8$ in RBF channel.}
        \label{fig:aoudia_output_08x04}
    \end{figure}

    A comparison of the performance of models trained using the model-aware technique
    using explicit channel model
    (trained using Adam, with knowledge of channel function $h(\cdot)$ and $\bm{\theta}_C$) 
    and model-free technique proposed in \cite{aoudia2018model} with objective
    function (\ref{exp:obj_rbf_1hot}) is given in Fig. \ref{fig:aoudia_output_08x04}.
    Here, $\sigma$ is the standard deviation of the Gaussian perturbation applied at
    the transmitter output.
    We can observe that the BLER performance (Fig. \ref{fig:aoudia_output_rbf_best_bler_08x04}) 
    of models trained using \cite{aoudia2018model} is 
    almost the same as the models which require channel knowledge (trained using Adam).
    However, in the case of packing density (Fig. \ref{fig:aoudia_output_rbf_en_08x04}), 
    the models trained with \cite{aoudia2018model} is slightly worse than models 
    trained using Adam. This could be explained as the added perturbation also acts 
    a noise to the model. The only price we pay while using \cite{aoudia2018model}
    to train is the slow convergence of the models.

	\bibliographystyle{IEEEtran}
    \bibliography{library.bib}
\end{document}